\documentclass{ws-rv9x6}
\usepackage{subfigure}   
\usepackage{ws-rv-thm}   
\usepackage{ws-rv-van}   

\usepackage[greek,english]{babel}
\usepackage{upgreek}
\usepackage{orcidlink}
\usepackage{mathrsfs}
\makeindex

\def\be{\begin{equation}}
\def\ee{\end{equation}}
\def\bea{\begin{eqnarray}}
\def\eea{\end{eqnarray}}
\def\bpar{\left(\!\!\begin{array}}
\def\epar{\end{array}\!\!\right)}
\def\bdar{\left|\!\!\begin{array}}
\def\edar{\end{array}\!\!\right|}
\def\barr{\begin{array}}
\def\earr{\end{array}}
\def\btab{\begin{tabular}}
\def\etab{\end{tabular}}
\def\nnb{\nonumber}
\def\vac#1{{\bf #1}}
\def\bnabla{\pmb\nabla}
\def\<{\langle}
\def\>{\rangle}

\def\half{{\textstyle\frac 12}}
\def\quarter{{\textstyle\frac 14}}
\def\MFT{{\scriptscriptstyle\rm MFT}}
\def\G{{\scriptscriptstyle\rm G}}

\definecolor{antiquefuchsia}{rgb}{0.57, 0.36, 0.51}
\definecolor{burgundy}{rgb}{0.5, 0.0, 0.13}

\definecolor{MyDarkGreen}{rgb}{0.02,0.60,0.06}
\definecolor{MyDarkGreen}{rgb}{1.00,0.00,0.00}

\def\mathcoppa{\hbox{\foreignlanguage{greek}{\coppa}}}

\def\qq{{\hbox{\foreignlanguage{greek}{\coppa}}}}
\def\qqq{{\hbox{\foreignlanguage{greek}{\footnotesize\coppa}}}}
\def\qqqq{{\hbox{\foreignlanguage{greek}{\tiny\coppa}}}}

\def\q{\hbox{\foreignlanguage{greek}{\coppa}}}
\def\qq{{\hbox{\foreignlanguage{greek}{\footnotesize\coppa}}}}

\def\coppa{{\fontencoding{LGR}\fontfamily{cmr}\selectfont\textqoppa}}
\def\q{\hbox{\foreignlanguage{greek}{\coppa}}}
\def\qq{{\hbox{\foreignlanguage{greek}{\footnotesize\coppa}}}}
\def\qq{\hbox{\fontencoding{LGR}\fontfamily{mtr}\selectfont\foreignlanguage{greek}{\coppa}}}

\begin{document}

\chapter[The enigmatic  exponent \coppa~and the story of finite-size scaling above $d_{\rm uc}$]{The enigmatic  exponent \coppa~and the story of finite-size scaling above the upper critical dimension\label{ra_ch1}}

\author[R. Kenna and B. Berche]{Ralph Kenna\orcidlink{0000-0001-9990-4277} and Bertrand Berche\orcidlink{0000-0002-4254-807X}\footnote{bertrand.berche@univ-lorraine.fr}}

\address{Centre for Fluid and Complex Systems,\\ 
Coventry University, Coventry, CV1 5FB, United Kingdom \\
Laboratoire de Physique et Chimie Th\'eoriques,\\ 
Universit\'e de Lorraine - CNRS, UMR 7019, Nancy,
B.P. 70239, F-54506 Vand\oe uvre les Nancy, France\\
${\mathbb L}^4$ Collaboration \& Doctoral College for the Statistical Physics of Complex Systems, \\ Leipzig-Lorraine-Lviv-Coventry, Europe\\
}

\begin{abstract}
Scaling, hyperscaling and finite-size scaling were long considered problematic in theories of critical phenomena in high dimensions. 
The scaling relations themselves form a model-independent structure that any model-specific theory must adhere to, and they are accounted for by the simple principle of homogeneity.
Finite-size scaling is similarly founded on the fundamental idea that only two length scales enter the game --- namely system length and correlation length.
While all scaling relations are quite satisfactory for multitudes of physical systems in low dimensions, one fails in high dimensions.
The aberrant scaling relation is called hyperscaling and involves dimensionality itself. 
Finite-size scaling also appears to fail in high dimensions.
Developed in the 1930s, Landau mean-field theory is valid in such high-dimensional systems. 
However, it too does not accord with hyperscaling and finite-size scaling there.
The advent of renormalization-group theory in the 1970s brought deeper fundamental insights into critical phenomena, allowing systems to be viewed at different scales. 
Above a critical dimensionality, higher-order Renormalization Group (RG) eigenvalues become irrelevant and scaling is governed by the Gaussian fixed point.
Although obeying all scaling relations including hyperscaling, and although it appears to successfully explain scaling in the correlation sector, the Gaussian fixed point fails to capture the free energy and derivatives, even in infinite volume.
In the 1980s, to fix this for the magnetisation, specific heat and susceptibility, Fisher introduced the notion of dangerous irrelevant variables. 
Since the correlation sector did not appear to be broken, no attempt was made to repair it and Fisher's modified RG formalism worked quite well in the thermodynamic limit of infinite volume.
However, finite-size scaling fails.
Also in the 1980's, Binder, Nauenberg, Privman and Young extended Fisher's concept to the free energy itself and to finite-size systems.
While putting Fisher's ideas on a more fundamental footing, the failure of finite-size scaling there still presented a problem.
This appeared to be resolved by the introduction of ``thermodynamic length'' to replace correlation length as the length scale that controls Finite-Size Scaling (FSS).
Thus hyperscaling and FSS were both sacrificed in favour of the RG patched together by ad hoc solutions.
In the 1990's,  Luijten and Bl\"ote went a long way to resolving the dilemma by adding corrections to scaling to the above considerations.
It was clear that both of these played a role.
However, as with previous authors, and adhering to the principle of not fixing that appears not to be broken, they did not address correlation length directly.

Here we report on developments over the past decade which went a long way to addressing these long-standing problems
The key to unlocking these, and extending their validity to the high-dimensional regime, was to relax assumptions that the correlation length has to be bounded by the physical length of bounded systems. 
This allowed and necessitated the extension of Fisher's concept of dangerous irrelevant variables to the correlation sector. 

\end{abstract}


\body

\tableofcontents

\section*{To the memory of my friend, Ralph Kenna}
Ralph and I started writing this review article for a volume of the series edited by our common friend Yurij Holovatch while Ralph was already ill. He knew that he would not see the book published. We had worked a bit earlier on a lecture format published in Scipost~\cite{ourscipost}, and this was not an easy task to propose a different perspective. This is Ralph's idea to write a kind of story, with many personal views and quotations.
The material presented here covers about fifteen years of our collaboration, and there is a lot to say. The result is maybe not really a scientific paper in the usual sense. As I wrote above this is more like a story,  as the title says already. Unfortunately, Ralph passed away before this was finished. The reader will easily recognize which parts were written by Ralph (the good ones), and which are mine (those with little originality). 

Those who met Ralph knew that he was a wonderful person. He was always enthusiastic, he was bringing new ideas onto the scene and was ready to fight, in a positive sense, in support of his opinions. He has been an excellent friend and an outstanding collaborator. We miss him. 

 This article is for a volume edited by our Ukrainian friend, (YuH). At the time I am finishing it,  after more than two years of war imposed on his country by Putin's government of Russia, another terrible war, imposed by Netanyahu's government of Israel, is devastating the Gaza strip, denying the right of Palestinian people to live there. Ralph is no longer there to tell me how he would feel bad to witness such a horrible situation in the Middle East, but I know how he, like many in Ireland, was against any kind of apartheid and colonisation. This is a hard time for the world when international laws and international regulations of Human Rights are trampled in the complete disinterest of most countries.  Many of these countries claim at the same time that they are democracies and claim that they work for Human Rights. Those countries that sell weapons to destroy Gaza. As simple human beings, we are against any kind of discrimination, any kind of racism, antisemitism, or islamophobia. Today my thoughts are in support of the Palestinian people who are suffering a tragedy imposed by Netanyahu's government of Israel. History and the international courts will qualify these mass killings as they should be.  Our thoughts have been on Israeli victims of the hideous attacks of the 7th of October, today, they are for Palestinian children\footnote{\href{https://www.amnesty.org/en/latest/news/2024/02/israel-opt-new-evidence-of-unlawful-israeli-attacks-in-gaza-causing-mass-civilian-casualties-amid-real-risk-of-genocide/}{https://www.amnesty.org/en/latest/news/2024/02/israel-opt-new-evidence-of-unlawful-israeli-attacks-in-gaza-causing-mass-civilian-casualties-amid-real-risk-of-genocide/}}. 
\index{Palestine}\index{Ukraine}

\hfill April 2024

\section{Introduction }
\label{SecNotations}

In the proceedings of a conference held in Washington DC in April 1965, Michael Fisher
\index{Fisher, M.E.} published what is an important but widely forgotten paper titled ``Notes, definitions, and formulas for critical point  singularities''~\cite{FisherHistorical}.
Prior to Fisher's intervention, different people used different symbols at different times for different measures.
Fisher pointed to us this when he invited us to meet him at the Royal Society in London, just before his retirement in October 2012. 
The paper opens with the introduction:
\begin{quote}
    ``For the convenience of participants a set of definitions and notes concerning the various critical point singularities and a table of formulas relating the lattice gas and Ising ferromagnetic models were distributed at the Conference. These are reproduced here with a few corrections and extensions.''
\end{quote}
This is how the worldwide practice of using  $\alpha$, $\beta$, $\gamma$, $\delta$, $\eta$ and $\nu$ (arranged in that order in Fisher's paper) was introduced and accepted in the community for critical exponents of the specific heat,  spontaneous magnetisation, susceptibility, induced magnetisation, anomalous dimension and correlation length. 
\index{critical!exponent}\index{exponent!critical}\index{critical!point}
Of course, phase transitions can only happen for infinite-sized systems and there is no mention of finite volume in Fisher's paper; the term ``critical point singularities'' in the title makes that clear.

In his paper, Fisher defined a generic critical exponent  $\lambda$ in the following way: 
\begin{quote}
   `` if
    \[
    \lim_{x\rightarrow 0^+}{\log{f(x)}}/\log{x} = \lambda\,
    \]
    we may say 
    \[
     f(x) \sim x^{\lambda}~\mathrm{as}~x\rightarrow 0^+.
     \]
     Note that with this definition the statement
     $f(x) \sim x^\lambda $ 
      does not exclude the possibility of a logarithmically divergent factor, i.e., 
      $Ax^\lambda |\log{x}| \sim  x^\lambda$. 
      In particular if $\lambda = 0$ the function $f(x)$ might diverge as $|\log{x}|$, for example, or $f( 0 )$ might be finite, the function $f( x)$ then being either continuous or discontinuous as $x$ passes through zero.''
\end{quote}
This is an example of the meticulous attention to detail characteristic of Fisher; he had not forgotten logarithmic corrections as an essential feature in the physics of critical phenomena.  \index{correction!logarithmic}
Although explicitly mentioning logarithmic corrections to the specific heat when $\alpha$ is zero in the two-dimensional Ising model, for example, he did not address them further in his paper.
He did, however, distinguish between critical exponents at either side of the phase transition, using primed exponents for the broken phase and unprimed ones for the symmetric phase.
Thus  $\alpha $ was used as the critical exponent for the specific heat when temperature $T$ exceeds its critical value $T_c$, for example, while  $\alpha^\prime$ was used for $T<T_c$.

This is why when one of us reported on \index{logarithmic!exponent}
\index{logarithmic!correction}
the exponents of logarithmic corrections in  Volume III of this series (2012), hatted, rather than primed notations, were used~\cite{Kennalog}. 
Thus if  $f(x)$ has a logarithmic correction it is expressed as $f(x) \sim x^\lambda | \ln{x}|^{\hat{\lambda}}$ with $\hat{\alpha}$, $\hat{\beta}$, $\hat{\gamma}$, $\hat{\delta}$, $\hat{\eta}$ and $\hat{\nu}$ representing exponents of logarithmic corrections for the observables listed above.
Just as there are scaling relations between the leading critical exponents, so too are there relations between their logarithmic counterparts. 
Indeed, it was the introduction of these scaling relations in Refs.~[\refcite{KennaPRL1,KennaPRL2}] (and in Volume III of this series) that captured Fisher's attention and led to our memorable invitation to meet him at the Royal Society. 

However, the analogue is not perfect; for the scaling relations between the hatted exponents to match exact and numerical calculations for specific models, there needed to be a logarithmic correction to the finite-size behaviour of the correlation length.
This was termed  $\hat{q}$ in Refs.~[\refcite{Kennalog,KennaPRL1,KennaPRL2}]
and had no counterpart in Fisher's list. 
Indeed, it referred to finite systems where a singularity cannot emerge, so that if $L$ denotes the linear extent of a finite system, the correlation length scales as $ \xi_L \sim L (\ln{L})^{\hat{q}} $.
Serendipitously, the missing counterpart emerged the same year as our 
meeting with Fisher in our paper about hyperscaling above the upper critical dimension
published as~Ref.~[\refcite{BERCHE2012115}]. 
Since we had used $\hat{q}$ for the exponent governing logarithmic corrections to finite-size scaling [FSS] of the correlation length, it was natural to use  $q$ as its leading counterpart so that $ \xi_L \sim L^{q} $.\footnote{The symbol $\hat{q}$ was used in Ref.~[\refcite{KennaPRL1}] simply because $q$ is next in the alphabet after $p$, which was used instead of the gap exponent for scaling of Lee-Yang zeros in an earlier version of that paper. This usage of $p$ was intended to indicate that the theory set out in Ref.~[\refcite{KennaPRL1}] was not reliant on concepts such as renormalization-group eigenvalues. While the eventual usage of $\Delta$ instead of $p$ preempts the double meaning of ``gap'' (both meanings are explained below), the consequent usage of $q$ and $\hat{q}$ was serendipitous,  as it fits very well with the convolution of related literature, as we shall see below.} 

``There is one thing wrong with your theory'' were the terrifying words of Michael Fisher at our meeting --- ``the letter `$q$' is not right''. 
It did not match the Greek alphabet! 
Our relief that the deficiency in our theory rested only with nomenclature, and not in bringing finite-size concepts into descriptions of (infinite-volume) critical-point singularities,
emboldened us to offer an excuse that all the Greek letters had been used up!
This was met with calm assurance  that again revealed Fisher's meticulousness
 as he etched out the symbol ``\coppa'': ``use ``coppa'', he said ``- its an archaic Greek letter''.
Thus a new symbol entered the pantheon tamed by Fisher nearly half a century earlier.
\index{Fisher, M.E.}

Our claims of a superlinear correlation length in high dimensions raised eyebrows in some quarters, as did the usage of \coppa.  \index{coppa}
Resistance to the former was due to dogmatic belief that the correlation length {{dare not}} exceed the length of a finite system, despite this already having been proved in specific models twenty years earlier by {\'{E}}douard Br{\'{e}}zin~\cite{Brezin82}.
Hostility to the nomenclature is captured in the statement: ``The authors’ introduction of their new exponent with the ankh notation is thus a gimmick which is not recommended in the scientific literature.
Instead, they should perhaps find a new letter that can be easily typeset.''
Faced with the options of defying that anonymous referee or Michael Fisher, we did not waver  and {\emph{A new critical exponent $\q$ and its logarithmic counterpart $\hat{\q}$}} was published in 
Condensed Matter Physics\footnote{The journal was founded in 1993 by the Institute for Condensed Matter Physics of the National Academy of Sciences of Ukraine. In March 2022 they issued the statement ``Our country has suffered an unjust and unprovoked aggression. Our army and our people are carrying an existential fight for Ukraine's very survival as an independent state, for our lives and freedom. Despite everything, we here, not at the frontline, are determined to do our job and continue publishing new issues of our journal on a regular basis, as long as it is possible. If you wish to support the Ukrainian scientific community, we are cordially inviting you to submit your manuscripts to our journal.'' We refer the reader to the journal's website: http://www.icmp.lviv.ua/journal}, John Cardy having come to the rescue for the typesetting \cite{CMP2013}.

Hyperscaling, which is one of the four scaling relations that interlink the six most prominent critical exponents \index{hyperscaling relation}
$\alpha$, $\beta$, $\gamma$, $\delta$, $\nu$ and $\eta$, usually reads as
\be
 \nu d = 2-\alpha .
 \label{hyperscaling1}
\ee
Here, $d$ is the dimensionality of the system and does not feature in the other three scaling relations (which we list in Subsection~\ref{homo}, below).
The  Ginzburg criterion (explained below) marks the critical value $d_{\rm uc}$ of dimensionality beyond which critical fluctuations of the order parameter cease to play a leading role.
Mean-field theory (MFT) is a valid first approximation there and the criterion dictates that $\nu$ and $\gamma$ are fixed to their critical-dimension values.
The remaining critical exponents, $\alpha$ included, are fixed  by the remaining scaling relations.
With $\alpha$ and $\nu$ so fixed, and $d$ free to roam above its critical value $d_{\rm uc}$  Eq(\ref{hyperscaling1}) cannot hold there and, for this reason, it is frequently said (including in textbooks, reviews and even new literature) that  FSS fails above the upper critical dimension.

The introduction~\cite{CMP2013} of the new exponent \coppa~renders such statements obsolete; a small alteration to the expressions (\ref{hyperscaling1})  based on a foundational alteration to the renormalization group, delivers a new form which is valid above (and below) $d_{\rm uc}$. 
In volume IV of this book series, we provided a renormalization-group basis for the correlation length $\xi_L$ to take the form 
\be
 \xi_L \sim L^{\qqq} ,
 \label{4}
\ee
with 
\be
\mathcoppa=\frac{d}{d_{\rm uc}}
\label{coppadoverdc1}
\ee
above $d_{\rm uc}$ and $\mathcoppa=1$ below.
The new hyperscaling relation is then \index{hyperscaling relation}
\index{critical!dimension!upper}
\be
 \nu \frac{d}{\mathcoppa} = 2-\alpha .
 \label{Q1}
\ee
This statement of hyperscaling is far from trivial - it is no mere ``gimmick'' or cosmetic alteration of  Eq.(\ref{hyperscaling1}) to phenomenologically render $\nu d_{\rm uc} = 2-\alpha$.
It is rooted in a very subtle and fundamental extension of the renormalization group to the correlation sector. 
The subtlety of  this extension is evinced by it having lain hidden for nearly fifty years --- precisely because of the success and simplicity of mean-field theory.

As stated, the superlinear behaviour of the correlation length was already introduced in Volume~III of this series in a renormalization-group-independent manner~\cite{Kennalog}. 
Ref.~\cite{Kennalog} reported on a series of papers \cite{KeLa91,KeLa93,KeLa93b,KeLa94,Ke2004}, where one of us presented a self-consistent theory of finite-size scaling at the upper critical dimension itself. 
There, logarithmic corrections modify mean-field behaviour in an essential way. 
Expressed into $\hat{\qq}$ instead of into $\hat{q}$,
the counterparts of Eq.(\ref{4}) and Eq.(\ref{coppadoverdc1})  are 
\be
 \xi_L \sim L (\ln{L})^{\hat{\qqq}} .
 \label{qr}
\ee
with 
\be
\hat{\q}=\frac{1}{d_{\rm uc}}.
\label{coppadoverdcc}
\ee \index{coppa}
These led to a logarithmic counterpart of the new hyperscaling relation (\ref{Q1}), namely \index{hyperscaling relation}
\be
 \hat{\nu} =  \hat{\qq} - \frac{\hat{\alpha}}{d_{\rm uc}} .
 \label{Q11}
\ee
One of the objectives of this chapter is to report on the origins of and necessity for the above formulae for hyperscaling  --- over fifty years after the old version was first set down~\cite{Wi65a,Wi65b,Gr67,Wi64}.
Eq.(\ref{coppadoverdcc}) holds in most cases but there are exceptions --- see Refs.~[\refcite{Kenna_2017,Juan}] for details. (We do not require those details for what is to come.)

\index{critical!dimension!upper}\index{finite-size scaling}
Finite-size scaling is also considered to fail above the upper critical dimension $d_{\rm uc}$.
Its standard form, valid below (but not above) the upper critical dimension, reads
\be
 \frac{Q_L(\tau=0)}{Q_\infty(\tau)} = \mathcal{F}_Q
 \left[{\frac{L}{\xi_\infty(\tau)} }\right].
 \label{FSS1}
\ee
Here $Q$ is a generic thermodynamic function such as specific heat $c$ or magnetic susceptibility $\chi$. 
The variable 
\be\tau=T/T_c-1\label{EQ_tauvsT}\ee 
represents the reduced temperature of the system and vanishes at the critical point of its infinite-volume thermodynamic limit.
The subscripts in Eq.(\ref{FSS1}) represent the system size and $\xi$ is the correlation length.
A hand-waving argument often used in favour of the FSS form (\ref{FSS1}) is that $L$ and $\xi$ are the only two length scales associated with a given system, so all forms of scaling should depend on their ratio.
If $Q_\infty(\tau)$ diverges as the critical point is approached as $|\tau|^{-\rho}$, say, this divergence has to be matched by the functional dependence of  $\mathcal{F}_Q$ on the correlation length. 
This delivers the classic FSS form \index{finite-size scaling}
\be 
 Q_L(0) \sim L^{\rho/\nu},
 \label{eq87}
 \ee
 which is not divergent for finite $L$.

In the case of magnetic susceptibility, this gives $\chi_L(0) \sim L^{\gamma / \nu}$. 
However, the FSS ratio $\gamma / \nu$ is not matched by Br{\'{e}}zin's exact calculations and numerical simulations by Kurt Binder for specific models above the upper critical dimension~\cite{Brezin82,Binder85}.
In particular, already in the 1980's both methods showed that $\xi_L \sim L^{5/2}$ for the 5D Ising model, rather than $L^2$ which is what standard FSS with $\gamma=1$ and $\nu=1/2$ would predict.   
Eq.(\ref{eq87}) therefore fails above the upper critical dimension and this is frequently stated in the same textbooks, reviews and other literature that condemn hyperscaling to failure there.
A primary outcome of our chapters in Volumes III and IV of this series
is the replacement of Eq.(\ref{FSS1}) by
\be
 \frac{Q_L(t=0)}{Q_\infty(t)} = \mathcal{F}_Q
 \left[{\frac{\xi_L(t=0)}{\xi_\infty(t)} }\right],
 \label{Q13}
\ee
where
\be
 t = \frac{T}{T_L}-1,\label{EQ_tvsT_L}
 \ee
in which $T_L$ is a pseudocritical point. \index{pseudocritical point}
We refer to this as QFSS to distinguish it from standard FSS.  \index{finite-size scaling}
There are two important changes in QFSS of Eq.(\ref{Q13}) relative to standard FSS in Eq.(\ref{FSS1}).
The first is the replacement  $\xi_L / \xi_\infty$ for $L / \xi_\infty$. 
The necessity for this change was already suggested in 1991 when, in Ref.~[\refcite{KeLa91}], it was used to theoretically identify and numerically test FSS for Fisher zeros and the specific heat for the $\phi^4$ model in four dimensions.
The second change is the replacement~\cite{BERCHE2012115} of reduced temperature $\tau=T/T_c-1$ by $t=T/T_L-1$ (see also Volume IV) \cite{doi:10.1142/9789814632683_0001}.

As stated, in Volume III, narrowly preceding our 2012 Royal-Society meeting, $\hat{q}$ was used rather than Fisher's $\hat{\q}$.
Three years later, in  Volume IV (2015), we presented a chapter on scaling above the upper critical dimension centered on $\phi^4$ theory and Ising models with short- and long-range interactions.
There the new archaic symbol $\q$ (and $\hat{\q}$) was deployed. 
In the meantime, the notation has been taken up in multiple studies of phase transitions in high dimensions, e.g. recently in Refs.~\cite{QuantumQFSS,adelhardt2024monte,PhysRevE.108.044146,bonamassa2024chimeric}.

Here, in  Volume VIII,  we again deploy the notation suggested by Fisher.
The purpose of this chapter is to generalise and extend some of the concepts presented previously. 
We start 
 by introducing very fundamental (model-independent)  Widom scaling concepts and extending them to the correlation sector when logarithmic corrections are present.
This latter move will be justified by posteriori, by a similar extension to the renormalization-group formalism.
These extensions are a main contribution of this body of work \cite{KeLa91,
KeLa93,
KeLa94,
KeLa93b,
Ke2004,
KennaPRL1,
KennaPRL2,
BERCHE2012115,
CMP2013,
flore2016,
PhysRevLett.116.115701,
doi:10.1142/9789814632683_0001,
Kennalog,
KB2014,
Kenna_2017,
ourscipost}.
On this basis, we re-derive three of the scaling relations for logarithmic corrections.
We also introduce the fundamental theory of phase transitions, finite-size scaling, and pseudocritical concepts. 
This provides a base for us to build upon Volumes III and extend Volume  IV to the more general case of $\phi^n$ theory.

\section{Homogeneity, scaling, logarithms and the enigmatic \coppa}
\label{SecNotations}
\index{homogeneity}\index{coppa}

\begin{quote}
``It should be clear that the purely thermodynamic approach employed 
in this paper cannot in and of itself tell us the behaviour of various 
functions near the critical point. 
Rather, it provides correlations among data obtained from experiments, 
or from statistical calculations, and checks on their consistency. 
The value of thermodynamics for checking consistency has already
been demonstrated.''   
\end{quote}
These are the words of Robert Griffiths whose name, through his paper \cite{Gr67}, is associated with one of the scaling laws for critical exponents. 
The same can be said for Refs.~[\refcite{KennaPRL1,KennaPRL2}] on which  Section~\ref{Sec3} of this chapter rests. 
The scaling theory presented there is based on self-consistencies, which are manifest as relations between the various exponents associated with logarithmic correction to 
scaling and finite-size scaling. 
For ab initio model-specific theories, the renormalization group 
and related approaches are appropriate.
That was the subject of our contribution to Volume~IV of this series.
There we addressed the $\phi^4$ model, a  model that trims back to the bare essentials of dimensionality and symmetry - the essence of universality.



In the usual spirit of the renormalization group, we consider a model with even and odd control parameters or fields $\tau$ and $h$. 
The even field $\tau$ may be thought of as a reduced temperature in the case of spin models:
$  \tau = T/T_c-1$.
The odd ordering field $h$ can be thought of as $h= \beta H$  where $\beta = 1/k_BT$ is the inverse temperature up to the Boltzmann constant $k_B$.
In Ising-type models, $H$ is the external magnetic field.
In the case of percolation theory, for example, where there is no temperature or magnetic field the probability of site occupation plays the role of $\beta$.

\subsection{Homogeneity}
\label{homo}
\index{homogeneity}\index{homogeneity!assumption}
A homogeneous function $f(\tau,h)$ is one that is only affected by a multiplicative rescaling if $\tau$ and $h$ themselves are multiplicatively rescaled. 
The Widom scaling hypothesis is that the singular part of the free energy density of $\phi^n$ theory is so governed  and
\be
f_\infty(\tau,h) = 
b^{-d}f_\infty( b^{{y_t}}  \tau, b^   {y_h} h),
\label{eq-10}
\ee 
where $b$ is an arbitrary rescaling factor~\cite{PatashinskiPokrovski,Ka66}. The precision that this is the singular part of the free energy density is important as we know that regular contributions do also exist but they do not enter the following discussion.

Here the subscript indicates we are dealing with a system of infinite extent in all $d$ directions. 
A similar hypothesis may be applied to the correlation function and the correlation length so that 
\bea
g_\infty(\vac x,  \tau,h)&=&b^{-2x_\phi}g_\infty(b^{-1} \vac x, b^   {y_t}  \tau, b^   {y_h} h) \label{eq-11}\\
\xi_\infty(  \tau,h)&=& b\ \!\xi_\infty( b^   {y_t}  \tau, b^   {y_h} h) . \label{eq-12}
\eea 
Here $\vac x$ is a 
measure of physical distance in any direction of the $d$-dimensional system and isotropy is assumed. 

If the theory exhibits a phase transition at $T=T_c$ or $\tau = 0$ and $H=h=0$ between a low-temperature phase which is ordered and a high-temperature one which is disordered, the two phases are distinguished by the value of an order parameter, and, at this critical point,  physically measurable properties become singular.
These thermodynamic quantities (magnetization, internal energy density, magnetic susceptibility and specific heat) are derivable from the free energy density as follows: 
\bea
&&m_\infty(\tau,h)=\frac{\partial f_\infty(\tau,h)}{\partial h} =b^{-d+   {y_h}}m_\infty( b^   {y_t}  \tau,b^   {y_h} h),\label{eq-1a}\\
&&e_\infty(\tau,h)=\frac{\partial f_\infty(\tau,h)}{\partial \tau} =b^{-d+   {y_t}}e_\infty( b^   {y_t}  \tau,b^   {y_h} h),\label{eq-1d} \\
&&\chi_\infty(\tau,h)=\frac{\partial^2 f_\infty(\tau,h)}{\partial h^2} = b^{-d+2   {y_h}}\chi_\infty(b^   {y_t}  \tau,b^   {y_h} h),\label{eq-1c}\\
&&c_\infty(\tau,h)=\frac{\partial^2 f_\infty(\tau,h)}{\partial \tau^2} =b^{-d+2   {y_t}}c_\infty(b^   {y_t}  \tau,b^   {y_h} h). \label{eq-1e}
\eea

In these relations, signs and multiplicative factors that do not compromise the form of the singularities are omitted.\index{singularities}

Thus the simple homogeneity assumption allows the various thermodynamic functions to be expressed in terms of only two scaling dimensions $y_t$ and $y_h$. 
We may express them in the conventional form that predates the renormalization group as follows~\cite{FisherHistorical}.
Firstly, setting $\tau=0$ and $b=|h|^{-1/   {y_h}}$,
\begin{eqnarray}
&m_\infty(0,h)\simeq D_{c}^{-1/\delta}|h|^{1/\delta}, 
    &\delta=\frac{   {y_h}}{d-   {y_h}},
    \quad D_{c}^{-1/\delta}=m_\infty(0,1).\label{defBc}
\end{eqnarray}
 Then, in zero magnetic field, setting $b=|  \tau|^{-1/   {y_t}}$ delivers\footnote{Repeated differentiation wrt field delivers higher moments so that the $n$th magnetic moment scales as $|\tau|^{(d-ny_h)/y_t}$. The ``gap'' between exponents is then $y_h/y_t=\beta+\gamma=\beta\delta$ and is conventionally given the label $\Delta$. }
\begin{eqnarray}\index{critical!amplitude}
&m_\infty(  \tau,0)\simeq B^- |\tau|^\beta, 
    &\beta=\frac{d-   {y_h}}{   {y_t}}, 
    \quad B^-= m_\infty(-1,0), 
    \    \tau < 0\, , \label{defB}\\
&\chi_\infty(  \tau,0)\simeq \Gamma^\pm|  \tau|^{-\gamma}, 
    &\gamma=\frac{2   {y_h}-d}{   {y_t}},
    \quad\Gamma^\pm = \chi_\infty(\pm 1,0),\label{defGamma}\\
&c_\infty(  \tau,0)\simeq\frac{A^\pm}\alpha|  \tau|^{-\alpha}, 
    &\alpha=\frac{2   {y_t}-d}{   {y_t}}, 
    \quad\frac{A^\pm}{\alpha}=c_\infty(\pm 1,0).\label{defA}
    \end{eqnarray} \index{critical!exponent}\index{exponent!critical}
    Here, we omit non-universal metric factors (see Refs.~\cite{PrivmanHohenbergAharony91,HattedScalingLaws,Berche_2013}).
We can apply a similar convention to the correlation function and correlation length to obtain
\be
 g_\infty(  \tau=0,h=0, \vac x)\sim\frac{1}{| \vac x|^{d-2+\eta}}, \label{eq-5} 
\ee
and
\bea
&\xi_\infty(  \tau,0)\simeq \Xi^\pm|  \tau|^{-\nu}, &\nu=\frac{1}{   {y_t}},\quad\Xi^\pm=\xi_\infty(\pm 1,0),\label{defNu}
\\
&\xi_\infty(  0,h)\simeq \tilde\Xi^\pm|  h|^{-\nu_c}, &\nu_c=\frac{1}{   {y_h}},\quad\tilde\Xi^\pm=\xi_\infty(0,\pm 1).\label{defNub}
\eea
Eliminating the scaling dimensions $y_t$ and $y_h$ in favour of the measurable critical exponent delivers
the scaling relations \index{scaling relations}
\bea
&\nu d=2-\alpha, 
\label{oldhyper} \\
&\alpha+2\beta+\gamma=2, 
 \label{alphabetagamma}\\
&\beta(\delta -1)=\gamma, \\
 \label{betadelta}
&\nu(2-\eta)=\gamma.
\label{oldeta} 
\eea

These famous scaling relations were developed in the 1960's through altogether different methods by Benjamin Widom\cite{Wi65a,Wi65b,Gr67,Wi64}, Leo Kadanoff\cite{Ka66}, 
John Essam and Michael Fisher\cite{EsFi63,Fi64sc}.
The first three are most commonly referred to as Josephson's~\cite{Jo67}, Rushbrooke's~\cite{Ru63} and Griffiths'~\cite{Gr65} scaling laws, in honour of their eponymous proponents, and the fourth was developed by Fisher~\cite{BuGu69,Fi69inequalities}.
The first and last scaling relations are conspicuous in that they involve the dimension of the system $d$ and the anomalous dimension $\eta$.
The first formula here recovers Eq.(\ref{hyperscaling1})
and, although named {\emph{hyperscaling}}, it does not hold for a hypercube such as the Ising model above four dimensions. 
As stated earlier, it only holds at and below the upper critical dimension.
The quantity $\eta$ in the last scaling relation is called the \emph{anomalous dimension} because its deviation from zero is a measure of deviation from mean-field  theory.
It governs the decay with distance of the  correlation function as
\index{anomalous dimension}
\be
d-2+\eta=2x_\phi.
\label{defEta}
\ee
David Nelson, in his remembrances\cite{DNelsonSays}, reports: \index{Fisher, M.E.}
\begin{quote}
``Fisher is uniquely responsible for the prediction of the anomalous critical exponent $\eta$ (Michael
once owned a boat that sailed on Lake Cayuga in Ithaca, New York, named the Eta.), which
controls the decay of order parameter correlations at the critical temperature. This exponent also
plays a key role in quantum field theories, where it is closely related to the anomalous scaling
dimensions. As Michael himself once said, the exponent is numerically small, but nevertheless
quite important.''
\end{quote}

We also find 
\be
 \nu_c=\frac{\delta-1}{2\delta}.
 \label{extraone}
\ee
The last formula is less commonly used and does not have a name (that we know of).
We display it distinctly here as it hides a subtlety emblematic of some of the features we wish to bring to the fore in this review.

To summarise, a simple application of the homogeneity assumption  (before the Renormalization Group (RG) comes into play) for the free energy, correlation function and correlation length, enables the critical exponents $\alpha$, $\beta$, $\gamma$, $\delta$, $\nu$, $\eta$, as well as $\nu_c$  to be written in terms of scaling dimensions $y_t$, $y_h$ and $x_\phi$ as well as the dimensionality $d$. 
This enables us to write down the scaling relations which were derived by other means in the 1960's (before the advent of the renormalization group).
We have not yet discussed the actual values of the scaling dimensions or critical exponents as these are model-specific.
We repeat that exact and numerical results show that hyperscaling in the form (\ref{oldhyper}) fails above the upper critical dimension. \index{critical!exponent}\index{exponent!critical}
Next, we do similar for logarithmic corrections - we refer to the literature for their independent derivation and re-derive them from the homogeneity assumption. 


\subsection{Logarithmic corrections}\index{correction!logarithmic}
\index{logarithmic!correction}\index{critical!dimension!upper}

At the critical dimension $d=d_{\rm uc}$, logarithmic corrections arise.
These were presented using a similar formalism to that used above for $\phi^4$ theory over 20 years ago by  Nevzat Aktekin~\cite{Aktekin2001}.
The introduction of logarithms, of course, loses the homogeneous properties of the free energy and related functions - multiplicative rescaling of the argument of a logarithm does not result in multiplicative rescaling of the logarithm itself! 
Notwithstanding this, inspired by the Privman-Fisher form (\ref{eq-10}) Aktekin proposed the following formula for the 4D Ising model ($n=4$):
\be
f^{\rm{s}}_L(\tau,h) =
L^{-4}
Y[{ L^{2} (\ln{L})^{1/6} \tau, 
                      L^{3} (\ln{L})^{1/4} h
                      }] \label{eq-10logA}.
\ee 
with $\tau = T/T_c-1$.
From this, FSS for the thermodynamic functions ensue, finding support in Ref.~[\refcite{Aktekin2001}] by Monte Carlo simulations on simple four-dimensional lattices linear extent up to $L= 16$ with periodic boundary conditions.
Aktekin justifies this structure by referring to previous work by Erik Luijten and Henk Bl\"ote~\cite{PhysRevB.56.8945}: 
``for the Ising model in $d=d_{\rm uc}$, the expression for $f^{\rm{s}}_L(\tau,h)$ derived starting with the renormalization group equations in differential form~\cite{PhysRevB.56.8945} reduces to the one given in Eq.(\ref{eq-10logA}) for $L \rightarrow \infty$ and confirms it.''
We return to Luijten and Bl\"ote's seminal contribution in Section~\ref{sec6.4} below.
 \index{finite-size scaling}

The contributions of homogeneous and inhomogeneous terms to the scaling formalisms of the renormalized Schwinger functions for the $\phi^4$ theory (including finite size) were also described in Refs.~[\refcite{BLZ,RGG,KeLa93}].
Below four dimensions, the inhomogeneous term is not divergent in the thermodynamic limit.
There the homogeneous term leads to critical behaviour (as captured in Subsection~\ref{homo} above). 
As we shall see,  the Gaussian fixed point is unstable for $d<4$ and does not govern large distance behaviour. 
Instead, the Wilson-Fisher fixed point governs critical behaviour.
In contrast, the Gaussian fixed point is stable above the critical dimension four.
As the critical dimension is approached, merged fixed point at the origin leads to a double zero of the so-called Callan–Symanzik beta function and this is responsible for the occurrence of logarithmic corrections.
The inhomogeneous term then contributes to the leading singular behaviour too. 
The homogeneous term remains singular, however, and, as shown in Ref.~[\refcite{KeLa93}], contributes to divergences such as that in the susceptibility.
Thus, while movement of the Wilson-Fisher fixed point to its Gaussian counterpart is responsible for triviality, the double zero gives rise to logarithmic corrections. 

Keeping focus on the inhomogeneous term (but not forgetting the well-known role still played by the homogeneous one as explicitly described in  Ref.~[\refcite{KeLa93}]), a straightforward extension to the general $\phi^n$ case was outlined in Ref.~[\refcite{Ke2004}].
Besides the trivial background term (\ref{eq-10}) (see Refs.~[\refcite{KeLa93}] and [\refcite{PhysRevB.56.8945}]),
one accounts for leading logs through~\cite{Kennalog}.

\be
f_\infty(\tau,h) =
b^{-d}
f_\infty[{ b^{y_t} (\ln{b})^{\hat{y}_t} \tau, 
                      b^{y_h} (\ln{b})^{\hat{y}_h} h
                      }] \label{eq-10log}.
\ee \index{homogeneity!assumption!with logarithms}
At the risk of stating the obvious, the same formula was recently claimed in Ref.~[\refcite{Deng2021}] with the explicit background term but without the backing of theoretical motivation of Luijten and Bl\"ote's RG calculations~\cite{PhysRevB.56.8945}.

Thermodynamic quantities are derivable as before but with logarithmic corrections:
\bea
m_\infty(\tau,h) &=& b^{-d+{y_h}}
(\ln{b})^{
 {\hat{y}_h}}m_\infty[{ b^   {y_t} (\ln{b})^{\hat{y}_t}  \tau,b^   {y_h} (\ln{b})^{\hat{y}_h}  h}],\label{eq-1alog}\\
e_\infty(\tau,h)&=& b^{-d+{y_t}}
(\ln{b})^{
 {\hat{y}_t}}e_\infty[{ b^   {y_t} (\ln{b})^{\hat{y}_t}  \tau,b^   {y_h}(\ln{b})^{\hat{y}_h}  h}],\label{eq-1dlog} \\
\chi_\infty(\tau,h)&=&  b^{-d+2{y_h}}
(\ln{b})^{2\hat{y}_h}\chi_\infty[{ b^   {y_t} (\ln{b})^{\hat{y}_t}  \tau,b^   {y_h} h (\ln{b})^{\hat{y}_h} }],\label{eq-1clog}\\
c_\infty(\tau,h)&=& b^{-d+2   {y_t}}
(\ln{b})^{2\hat{y}_t}c_\infty[{b^   {y_t} (\ln{b})^{\hat{y}_t}  \tau,b^   {y_h} h (\ln{b})^{\hat{y}_h} }]. \label{eq-1elog}
\eea
Setting
$b=   |h|^{-1/y_h} (\ln{|h|})^{-\hat{y}_h/y_h}$
or 
$b=|\tau|^{-1/y_t} (\ln{|\tau|})^{-\hat{y}_t/y_t}$, appropriately, delivers
\index{correction!logarithmic}
\begin{eqnarray}
m_\infty(\tau,0)    & \sim 
                 & {|\tau|}^{\beta}|\ln{|\tau|}|^{\hat{\beta}} \quad {\mbox{for $T<T_c$}},
                 \quad  \hat{\beta} = \beta \hat{y}_t+\hat{y}_h
                 \label{mtlog}\\
m_\infty(0,h)    & \sim
                 & {|h|}^{\frac{1}{\delta}}|\ln{|h|}|^{\hat{\delta}},  
                 \quad   \hat{\delta} = d\frac{\hat{y}_h}{y_h}
                 \label{mhlog}\\
\chi_\infty(\tau,0) & \sim 
                 & {|\tau|}^{-\gamma}|\ln{|\tau|}|^{\hat{\gamma}}, \quad  \hat{\gamma} = - \gamma \hat{y}_t + 2\hat{y}_h 
                 \label{chitlog}\\
c_\infty(\tau,0)    & \sim 
                 & {|\tau|}^{-\alpha}|\ln{|\tau|}|^{\hat{\alpha}},  
                 \quad  \hat{\alpha} = d\frac{\hat{y}_t}{y_t}
                 \label{ctlog}.
\end{eqnarray}
Note that all hatted exponents are by default defined with a positive sign, even when the leading singularity is diverging. \index{critical!exponent!hatted}\index{exponent!hatted}

A counterpart formula for the correlation length was not considered in Ref.~\cite{Aktekin2001}. 
In anticipation of superlinearity, we express it as
\be
\xi_\infty(  \tau,h) 
= 
 b\ \!(\ln{b})^{\hat{\qq}}
\xi_\infty [{ b^{y_t} (\ln{b})^{\hat{y}_t} \tau, 
                      b^{y_h} (\ln{b})^{\hat{y}_h} h
                      }] .
                      \label{coppahat} 
\ee    
Here we have used the notation $\hat{\q}$ in anticipation of what is to come
(as stated, until Ref.~[\refcite{CMP2013}], we had used the more prosaic  symbol $\hat{q}$).  We thus have now
\begin{eqnarray}
{\xi_\infty(\tau,0)}  & \sim 
                 & {|\tau|}^{-\nu}|\ln{|\tau|}|^{\hat{\nu}},               
                 \quad  \hat{\nu} = -\frac{\hat{y}_t}{y_t} + \hat{\q}
                 \label{xitlog}.
\end{eqnarray}

Eliminating the scaling fields $\hat{y}_t$ and $\hat{y}_h$ in favour of the measurable critical exponents,\index{critical!exponent}\index{exponent!critical}
\begin{equation}
 \hat{y}_t = \frac{\hat{\beta}}{\beta} - \frac{\delta \hat{\delta}}{\beta(1+\delta)}
          = \frac{2 \hat{\beta} - \hat{\gamma}}{2 \beta + \gamma}
\label{delivery1}
\end{equation}
and
\begin{equation}
 \hat{y}_h = \frac{\delta \hat{\delta}}{1+\delta}
          = \frac{\gamma \hat{\beta} + \beta \hat{\gamma}}{2 \beta + \gamma},
\label{delivery2}
\end{equation}
delivers
the scaling relations for logarithmic corrections
\bea
 \hat{\alpha} & = & 2 \hat{\beta} - \hat{\gamma},
 \label{alphahao} \\
 (\delta-1)\hat{\beta} & = & \delta \hat{\delta} - \hat{\gamma},
\label{betao} \\
 \hat{\nu} & = & \hat{\q} - \frac{\nu (2\hat{\beta} - \hat{\gamma})}{2-\alpha}.
 \label{enigmatic}
\eea
Eqs.(\ref{alphahao}) and (\ref{betao}) are logarithmic counterparts for the scaling relations (\ref{alphabetagamma}) and (\ref{betadelta}), respectively, and inserting them in
Eq.(\ref{enigmatic}) delivers Eq.(\ref{Q11}).
These scaling relations for logarithmic corrections were developed and verified in Refs.~[\refcite{KennaPRL1,KennaPRL2}] through very different means (not reliant on phenomenologically modified Widom scaling).


In Ref.~[\refcite{KennaPRL1}], one of us stated ``relations (\ref{alphahao}) and (\ref{betao}) but not (\ref{enigmatic}) can be derived starting with a suitably modified phenomenological Widom ansatz.''
That ansatz was Eq.(\ref{eq-10log}),  later put to writing for the 4D Ising model by 
Aktekin in~Ref.~[\refcite{Aktekin2001}].
What was missing then was the extension of the ``suitably modified phenomenological Widom ansatz'' to the correlation sector, namely Eq.(\ref{coppahat}) and
the corresponding homogeneity assumption for the correlation function which reads as~\cite{HattedScalingLaws}
\be
g_\infty(\vac x,\tau,h)=
b^{-2x_\phi} (\ln{b})^{\hat{\eta}}
g_\infty [b^{-1}\vac x,{ b^{y_t} (\ln{b})^{\hat{y}_t} \tau, 
                      b^{y_h} (\ln{b})^{\hat{y}_h} h
                      }] .
                      \label{ghat} 
\ee

That extension, together with Eq.(\ref{coppahat}), had no basis then because of the evidence that was lacking. 
Evidence comes from two sources. 
The first basis is empirical; as documented in Volume~III of this series~\cite{Kennalog}, the three formulae (\ref{alphahao}), (\ref{betao}) and (\ref{enigmatic}) hold for different universality classes, including the Ising model, $O(N)$ $\phi^4$ models, long-range Ising models, $m$-component spin glasses, percolation, the Yang-Lee edge and lattice animals, all at their respective upper critical dimensions.
In each of these cases if  $\hat{\q}$ were set to zero the scaling relation would fail. 

The second basis for Eq.(\ref{coppahat}) was reported on in Volume~IV, some of which we next elaborate upon.

Let us mention here that in Ref.~\cite{Kennalog}, one of the scaling laws among hatted exponents, $\hat\eta=\hat\gamma-\hat\nu(2-\eta)$, was not correct in the case $\hat\qq\not=0$. Indeed, we reported in Ref.~\cite{HattedScalingLaws} the correct form, \be\hat\eta=\hat\gamma-\hat\nu(2-\eta)+\gamma\frac{\hat\qqq}{\nu}\ee
satisfying the condition
$g_L (\tau,h,|\vac r|\to\infty)\to m_L^2(\tau,h)$ which was erroneously relaxed in 
Ref.~\cite{Kennalog}.

\subsection{The enigmatic nature of  \texorpdfstring{$\hat{\qq}$}{TEXT} and of \texorpdfstring{$\qq$}{TEXT} }

To enable a physical interpretation of $\hat{\q}$, \index{coppa!hat}
we escalate Eqs.(\ref{eq-10log}), (\ref{coppahat}) and (\ref{ghat}) to allow for finite size:
\bea
f_L(\tau,h,L^{-1}) 
& = &
b^{-d}
f_L[{ b^{y_t} (\ln{b})^{\hat{y}_t} \tau, 
                      b^{y_h} (\ln{b})^{\hat{y}_h} h, 
                      bL^{-1}
                      }] \label{eq-10logL},
                      \\
\xi_L( \tau,h,L^{-1}) 
& = & 
 b\ \!(\ln{b})^{\hat{\qq}}
\xi_L[{ b^{y_t} (\ln{b})^{\hat{y}_t} \tau, 
                      b^{y_h} (\ln{b})^{\hat{y}_h} h,
                      bL^{-1}
                      }] ,
                      \label{coppahatL} 
                      \\
g_L( \vac x,\tau,h,L^{-1}) 
& = & 
b^{-2x_\phi} (\ln{b})^{\hat{\eta}}
g_L[{b^{-1}\vac x, b^{y_t} (\ln{b})^{\hat{y}_t} \tau, 
                      b^{y_h} (\ln{b})^{\hat{y}_h} h,
                      bL^{-1}
                      }]. \nonumber\\
                      \label{ghatL}                       
\eea 
As described in Ref.~[\refcite{KeLa93}], this is possible in RG terms because of its local nature --- the renormalization constants that apply in the infinite-volume theory, apply for finite volume  too~\cite{KeLa93}.
Setting $\tau=h=0$ in the first two arguments of $\xi$, and $b=L$ in the third argument (to sit at or in the vicinity of the pseudocritical point $t=0$),  we obtain Eq.(\ref{qr}) as the logarithmic counterpart of Eq.(\ref{4}).
It also enables the QFSS form (\ref{Q13}) by expressing the free energy and its derivatives as a function of $\xi_\infty$ and $\xi_L$ instead of $\tau$ and $L$.

Despite the previous widespread dogma that the correlation length cannot exceed the length, this logarithmic superlinearity was explicitly derived by Br{\'{e}}zin
 for the large $N$ limit of the $N$-vector model (spherical model) for periodic boundary conditions (PBCs).  Br{\'{e}}zin expected that the  picture  ``is qualitatively unchanged for finite $N$''~\cite{Brezin82}.
In particular, the value $\hat{\q}=1/4$ was explicitly laid out there.
This was implicitly verified for PBCs in the four-dimensional Ising model Ref.~[\refcite{KeLa91}] and explicitly (also for PBCs) in Ref.~[\refcite{PhysRevB.71.174438}].  

The value $\hat{\q}=1/4$ extends to all values of $N$ in the  $O(N)$ theory. 
The $N$-dependency of partition function zeros and thermodynamic functions were also studied in Ref.~[\refcite{Ke2004}].
There it was found that leading logarithmic corrections to the finite-size dependency are independent of $N$ in the odd sector --- i.e., for Lee-Yang zeros, magnetic susceptibility and related functions.
In contrast, Fisher zeros, specific heat and other quantities associated with the even sector were found to be $N$-dependent. 
It is now well established on model-specific theoretic and scaling grounds, as well as from numerical simulations that the correlation length of a confined system can exceed its physical length.

And herein lies the enigmatic nature of $\hat{\q}$.
The three scaling relations (\ref{alphahao}), (\ref{betao}) and (\ref{enigmatic}) are valid for infinite size but one of them is sourced in finite-size concepts.
If hyperscaling (\ref{hyperscaling1}) is a scaling relation between critical exponents $\alpha$ and $\nu$, its logarithmic counterpart (\ref{Q11}) is too and $\hat{\q}$ is a critical exponent, just as $\hat{\alpha}$ and $\hat{\nu}$ are. 
On the other hand $\hat{\q}$ is a manifestation of finite-size; it characterises the correlation length through Eq.(\ref{coppahatL}) and does not appear in infinite volume systems where only the exponent $\nu$ describes the correlation length in Eq.(\ref{defNu}).
In that case, $\hat{\q}$ is a pseudocritical exponent and Eq.(\ref{Q11}) is a finite-size relation.
But what then is the status of Eq.(\ref{hyperscaling1}) if it is to be replaced by Eq.(\ref{Q1})? 

In anticipation of a RG foundation, we next relax 
Eq.(\ref{eq-12}) to encompass superlinearity
as we have done for 
Eq.(\ref{coppahat})
\be
 \xi_\infty(  \tau,h) 
  =  
b^{\qq}
\xi_\infty[{ b^{y_t}  \tau, 
                      b^{y_h} h
                      }]. 
                      \label{coppainWidom} 
\ee 
With $\q=d/d_{\rm uc}$ as in Eq.(\ref{coppadoverdc1}) this delivers the modified hyperscaling form (\ref{Q1}).
Eq.(\ref{coppadoverdc1}) itself was also derived by   Br{\'{e}}zin for PBCs in Ref.~[\refcite{Brezin82}] and was 
 numerically verified by Jeff  Jones and Peter (A.P.) Young in Ref.~[\refcite{PhysRevB.71.174438}].  
There it was stated that ``for free boundary conditions, it seems obvious that even for $d > 4$ the behaviour of the system will be affected when $\xi_L$ becomes of order $L$, rather than only change when $\xi_L$ becomes of order the much larger length [$L^{5/4}$]''~\cite{PhysRevB.71.174438}.
The expectation in Ref.~[\refcite{PhysRevB.71.174438}] was that 
standard  FSS should apply for free boundary conditions (FBCs) and that the ratio $\xi_\infty/\xi_L$ ``may also enter'' but presumably for ``corrections to the scaling terms which involve $\xi_\infty/L$''.
Ref~\cite{PhysRevB.71.174438} ends with the statement ``Since FSS for models with FBCs in $d > 4$ is poorly understood, it would be interesting to investigate such models in some
detail.''
 
Thus we arrive at a rather puzzling situation.
On the one hand, if the new hyperscaling from (\ref{Q1}) and its logarithmic counterpart (\ref{Q11}) are to stand as scaling relations, $\q$ and $\hat{\q}$ would have to be considered as scaling exponents. 
But since their physical interpretations are finite-size, they are pseudocritical (like $\lambda$ and $\rho$ to be discussed below).
Moreover, since the critical exponents listed by Fisher in Ref.~[\refcite{FisherHistorical}] are associated with singularities, 
they are not burdened by limitations of boundary conditions. 
So the question arises to what extent  $\q$ and $\hat{\q}$ are universal - independent of boundaries. \index{coppa}
This is the main topic of this chapter.

\subsection{Summary }

At this point, we have  introduced a very general scaling picture based on a simple assumption of homogeneity in the free energy, correlation function and correlation length.
This delivers the standard scaling relations which have been in place since the 1960's and are known to work for $d \le d_{\rm uc}$.
We then dropped homogeneity and introduced logarithmic corrections to this picture with hatted exponents for each thermodynamic function.
In anticipation of the next section, this included an exponent $\hat{\q}$ which has no leading-scaling counterpart below the critical dimension.
To give physical meaning to  $\hat{\q}$  we promoted the logarithmically-corrected  form to finite-size systems. 
Thus captured the QFSS form (\ref{Q13}) for the critical dimension but not above it. 
Moreover, at the critical dimension itself, this superlinear scaling behaviour has mathematical and numerical backup, at least for the Ising model and for PBCs, and by extension for the $O(N)$ model, at least in the magnetic (odd) sector. 
At this point, then, the situation for FBCs is ``poorly understood'' and puzzling.
Universality suggests $\hat{\q}$ should manifest there but conservative thinking (that the correlation length cannot exceed the length) suggests it cannot.

We have not, so far, paid attention to the symmetry group of a given underlying Hamiltonian or the structure of the Hamiltonian itself. 
We have only considered measurable parameters $\tau$ or $t$ and $h$ as driving the phase transition.
Next, we explore these concepts further as we allow them to take complex values!
To do this, we briefly revisit the origins of scaling relations in its broadest context as reported in Volume III~\cite{Kennalog}.
Our purpose here is not to revisit old ground covered in Volume III but to provide context for higher dimensions.
For the question arises what is the high dimensional counterpart of $\hat{\q}$ and Eq.(\ref{coppadoverdcc})\footnote{We mention here an early audacious hypothesis which has not been published ten years ago but was recently proposed to a special issue of the journal Condensed Matter Physics in memory of Ralph~\cite{kenna2023previously}.}.
This will lead us to the new hyperscaling relation Eq.(\ref{Q1}). 
Structuring the RG to achieve this relation opens a whole new area in high dimensional that has been hidden since Fisher first introduced dangerous irrelevant variables (DIVs).
\index{dangerous irrelevant variable}

\section{The fundamental theory of phase transition as a framework} 
\label{Sec3}

In addition to thermodynamic and correlation functions, advocates of the partition-function-zero approach have an extra tool at their disposal for the understanding of critical phenomena. 
One of these was Fa Yueh (Fred) Wu who described them as delivering a  ``fundamental theory of phase transitions" \cite{Wu}.
The zeros approach itself was invented in the 1950's (well before the discovery of the renormalization group) in two papers by  Tsung-Dao (T.D.) Lee  and Chen Ning Yang~\cite{PhysRev.87.404,PhysRev.87.410}.\footnote{A few years later, in 1957, this dynamic pair were awarded the Nobel Prize  ``for their penetrating investigation of the so-called parity laws which has led to important discoveries regarding the elementary particles.''}
As stated by Wu, ``these two papers have profoundly influenced modern-day statistical mechanics.''

Partition function zeros are not simple derivatives of the free energy and are usually considered on a model-by-model basis. \index{zeros!of the partition function}
In terms of the free energy, the partition function itself is given by 
\be
 Z_L(\tau,h) = \exp{\left[{-\frac{1}{k_BT} F_L(\tau,h)}\right]} .
\ee
This is a real function which cannot vanish for finite $F_L(\tau,h)$.
When the free energy becomes singular in the infinite-volume limit, however, it can.
Yang and Lee's insight was to relax usual (physical) restriction that the external field $h$ be a real parameter. 
As a complex variable it ``opened a new window''~\cite{Wu} by enabling the vanishing of the partition function at values of field now called Lee-Yang zeros.
The celebrated Lee-Yang circle theorem, which holds for models with certain symmetries such as the Ising model, states that these zeros occur when $h$ is purely imaginary [on a unit circle in the fugacity $\exp{(\beta h)}$ plane]. 
Symmetry demands this is the case whether the system is infinite in size or finite. 
While interesting, \index{zeros!Lee-Yang} \index{zeros!Fisher}
the  Lee-Yang  theorem is not required for the considerations that follow in this chapter.
Still, whatever its shape, the line of Lee-Yang zeros is known as the singular line.
In the symmetric phase, for temperatures above the critical one, this line stays away from the real magnetic-field $h$ axis and terminates at what is known as the Yang-Lee edge (this name, too, was proposed by Michael Fisher~\cite{Fi78}). 
As the (real) temperature drops to the critical value, the Yang-Lee edge approaches the real $h$ axis at its critical value $h=0$. 
At the critical temperature itself (in infinite volume) the line of zeros pinches the real axis at this point (from the upper and lower halves of the complex $h$ plane) and 
analytic continuation  from ${\rm{Re}}(h)>0$ to ${\rm{Re}}(h)<0$ is forbidden.
Hence a first-order phase transition is precipitated at $h=0$. 
We denote the Yang-Lee edge (in infinite volume) by $r_{\rm{YL}}(t)$.
Its approach to the real axis is  characterised by a power law with the possibility of logarithmic corrections~\cite{KeLa91}:
\begin{equation}
 r_{\rm{YL}}(\tau)  \sim |\tau|^{\Delta} |\ln{|\tau|}|^{\hat{\Delta}} ,
 \label{YL}
\end{equation}
where $\Delta$ is the gap exponent.\footnote{As stated earlier,  $\Delta$ represents the ``gap'' between critical exponents labelling sequential magnetic moments.
That it also describes the Yang-Lee ``gap'' is a fortuitous etymological co-occurrence.}
We may identify this, and its logarithmic counterpart, from the form of the free energy in Eq.(\ref{eq-10log}) which, on setting the first argument to a constant is a function of
$h |\tau|^{y_h/y_t} |\ln{|\tau|}|^{y_h\hat{y}_t/y_t-\hat{y}_h}$.
Since the partition function holds this functional dependency, its zeros occur at given values of $h$, including at $h=r_{\rm{YL}}(\tau)$,   with
\begin{equation}
 \Delta = \frac{y_h}{y_t}  =  \beta+\gamma = \beta \delta = \frac{\gamma \delta}{\delta - 1}
\label{edge}
\end{equation}
and
\begin{equation}
 \hat{\Delta} = \Delta \hat{y}_t - \hat{y}_h  
 = \hat{\beta} - \hat{\gamma}
.
\label{edge2}
\end{equation}

One may similarly investigate zeros in the complex temperature plane.
These were introduced in 1964 by Michael Fisher in a series of lecture notes \cite{FisherZeros} and are called Fisher zeros. \index{Fisher, M.E.}
While Lee-Yang zeros are those of the grand canonical partition function, the Fisher zeros are of the canonical partition function only.
As for Lee-Yang zeros, these remain away from the real (temperature) axis if the external magnetic field is non-critical, i.e., non-zero.
As $h$ approaches zero, the zeros approach the real axis and pinching occurs.
However, Fisher zeros are rarely considered in non-vanishing $h$.
Like Lee-Yang zeros in the complex field plane, Fisher zeros sometimes lie on curves in the complex temperature plane.
However, their distribution is more elaborate than a simple line and may even disperse across 2-dimensional areas (this mostly occurs in circumstances of anisotropy).
As for Lee-Yang zeros, we do not require the precise shape of Fisher zeros for most of the considerations that follow in this chapter (it was required for Volume III which concerned logarithmic corrections at low dimensions as well).

\subsection{Lee-Yang zeros}

\index{zeros!Lee-Yang}
Be it for Lee-Yang or Fisher zeros, for finite-size systems the singular line lacks continuity of its infinite-volume counterpart. 
Instead one has a discrete bead of zeros. 
This is clear by writing the partition function 
as a sum over all possible configurations of the microscopic degrees of freedom.
We can gather these configurations into a histogram of frequencies of even and odd macrostates - energy and induced magnetisation, respectively. 
For the Ising model, with spins $s_i$ at the sites $i$ of a regular lattice comprising $L^d$ nodes and $dL^d$ links, for example, these are 
$E=-J\sum_{(i,j)} s_is_j$  and 
$M=\sum_i s_i$, respectively. 
We set the strength of interaction $J$ to one so that $E$ can take integer values ranging from $-dL^d$ and $dL^d$ and $M= \sum_is_i$ takes integer values from $-L^d$ to $L^d$.
The grand canonical partition function  in
the presence of an external magnetic field $H$ is then
\begin{equation}
 Z_L(\beta,h)=\frac{1}{\cal{N}} 
\sum_{\{s_i\}} e^{-\beta 
 \left( 
       -E-HM
 \right)}
 =
 \frac{1}{\cal{N}} 
\sum_{\{s_i\}} e^{ 
 \left( 
       \beta E + h M
 \right)}
 \quad,
\label{gndcanIsing}
\end{equation}
where $\beta = 1/{k_BT}$ is the inverse of the Boltzmann constant times the temperature and $h=\beta  H$ is the reduced external magnetic field. 
The factor $\cal{N}$ is introduced for normalization purposes and serves no active role in what follows.  The partition function may be expressed as
\begin{equation}
Z_L(\beta,h) = \sum_{M=-L^d}^{L^d} \sum_{E=-dL^d}^{dL^d}
  \rho_L(E,M) e^{\beta E + hM}  \quad,
\label{gndcanIsing1}
\end{equation} 
where 
the  spectral  density  $\rho_L(E,M)$  denotes  the  relative   weight  of 
configurations having given values of $E$ and $M$. $Z_L(\beta,h)$  can be 
written as a  polynomial in the fugacity $z$ defined by
\begin{equation}
 z = e^{-2h} \quad,
\label{paperIII2}
\end{equation}  
as
\begin{equation}
Z_L(\beta,h) = z^{-\frac{1}{2}L^d}
  \sum_{n=0}^{L^d} \rho_{L,\beta}(n) z^n  \quad,
\label{paperIII3}
\end{equation}  
in which
\begin{equation}
 \rho_{L,\beta} (n) =   \sum_{E=-dL^d}^{dL^d} 
 \rho_L(E,L^d-2n) e^{\beta E } 
\label{paperIII4}
\end{equation}  
is an integrated density.
That  the  partition  function  (\ref{paperIII3})  and  the  corresponding free energy (as given by its logarithm) are completely analytic for finite
$L$,  establishes that no phase transition can occur in a finite-size system. 
However, as $L$ is allowed to go to infinity phase transitions manifest themselves as points of non--analycity. 
For a given value of $L$ and $\beta$,  $Z_L$ has strictly complex roots $z_j$. 
Again, this is consistent with the fact that 
the free energy is analytic at any real value of the magnetic field $H$ for a finite system; the partition function has no zeros at real $H$. 
Complex zeros do, however, exist and we can write the partition in factorized form:
\be
Z_L(\beta,h)=A_L(\beta,h)\prod_{j=1}^{L^d}{(z-z_j)},
\label{e153}
\ee
with $A_L(\beta,h)$ a smooth non-vanishing function.
The Lee-Yang theorem states that the zeros lie on the unit circle in the plane of complex variable $z$, which means they lie on the imaginary $h$ axis~\cite{PhysRev.87.404,PhysRev.87.410}.
As stated, this theorem is not required in what follows.

 From conformal invariance, we can just as well operate in the complex $h$ plane and we write the $j$th zero there as $h_j(\tau)$ where it is to be understood that the actual value is $L$-dependent.
 One may then write the finite-size free energy as a sum over Lee-Yang zeros,
\be
 f_{L}(\tau,h)= A^\prime_L(\tau,h) L^{-d}\sum_j\ln(h-h_j(\tau)] ,
\ee
where $A_L(\tau,h)$ is a normalizing factor which plays no active role in what follows. 
(Even in the thermodynamic limit, it contributes only to the regular part of the free energy and its derivatives and not to critical behaviour.)
We therefore drop it from our considerations.
Differentiating twice wrt field $h$ then gives
\be
 \chi_L(\tau,h)=-L^{-d}\sum_j\frac {1}{[h-h_j(\tau)]^2}
\label{eq-158}
\ee
(having dropped a regular additive term).
For a finite system, the susceptibility would manifest a singularity if the magnetic field $h$ coincided with a complex Lee-Yang zero $h_j(\tau)$.
However, for purely classical systems, complex $h$ is not physical.
Only the thermodynamic limit enables the zero impact onto the real axis to precipitate a real phase transition.
In this sense complex zeros may be considered as ``proto-critical'' points\cite{Fi78} -- they have the potential to become critical points.

If we assume the critical behaviour is dominated by the first few zeros, or even just the first zero, one arrives at
\be
 \chi_L(0,0) \sim L^{-d} 
 h_1(0)^{-2}.
 \label{sleep}
 \ee
 At this point, we deploy the QFSS form (\ref{Q13}) to determine each side of the equation.
This amounts to the substitution 
\begin{equation}
     |\tau| \rightarrow 
      L^{\frac{-\qqqq}{\nu} }
      (\ln{L})^{-\frac{\hat{\qqqq}-\hat{\nu} }{\nu}}
      \label{FSSpre} 
\end{equation}
in the scaling expressions (\ref{chitlog}) for the susceptibility and (\ref{YL}) for the Yang-Lee edge to find
\be
\chi_L \sim L^{\frac{\qqqq \gamma}{\nu}}(\ln{L})^{\hat{\gamma}+\gamma\frac{\hat{\qqqq}-\hat{\nu} }{\nu}}
\ee
and
\be
 h_1 \sim L^{-\frac{\qqqq\Delta }{\nu}}(\ln{L})^{\hat{\Delta}-\Delta\frac{\hat{\qqqq}-\hat{\nu} }{\nu}}.
 \label{FSSofh1}
\ee
Matching the leading scaling behaviours of these two expressions in Eq.(\ref{sleep}) delivers 
\be
 \gamma = 2\Delta - \nu d_{\rm uc}
 \label{snooze}
 \ee
 which, since hyperscaling in the form (\ref{hyperscaling1}) always valid for $d=d_{\rm uc}$, gives $\Delta = \beta + \gamma$ which we already had in (\ref{edge}).
 This validates our assumption that the critical behaviour of susceptibility is dominated by the scaling of the first zero. \index{hyperscaling relation}
 Note also that $\q$, or dimensionality $d$, has dropped out of this equation, so it is valid in all dimensions.
 Note also that without the $\q$ exponent (or if $\q=1$), Eq.(\ref{snooze}) would read $ \gamma = 2\Delta - \nu d$. This is a standard form for hyperscaling and it fails for $d>d_{\rm uc}$.
 
 Applying the same considerations to logarithmic corrections gives
 \be
  2 \hat{\Delta} 
  = - \hat\gamma 
   + ( 2 \Delta-\gamma)\frac{\hat{\qqq}-\hat{\nu} }{\nu}.
 \ee
 With Eq.(\ref{snooze}), this gives
 \be
  2 \hat{\Delta} 
  = d_{\rm uc}(\hat{\qq} - \hat{\nu}) 
  - \hat{\gamma}.
  \ee
  Note that $\hat{\q}$ has not dropped out of this equation.
  Adhering to the dogma that the correlation length cannot exceed the length of the finite-size system is equivalent to setting $\hat{\q} = 0$.
   If we follow this route, we find the logarithmic correction counterpart of the gap exponents:
   $ 2\hat{\Delta} = - (d_{\rm uc} \hat{\nu} + \hat{\gamma})$.
   This equation, however, is incorrect.
   
   In the late 1980s and early 1990s, before CERN confirmed the existence of the Higgs boson\footnote{Ten years ago (4 July 2012), the ATLAS and CMS collaborations at the Large Hadron Collider(LHC)
   announced the detection of the Higgs boson, as  predicted by the Standard Model of particle physics, and one year later Fran\c cois Englert and Peter Higgs won the Nobel Prize, having predicted the existence of the particle with Robert Brout, decades earlier.}, the question of triviality of the $\phi^4$ model was very much in vogue and logarithmic corrections played a crucial role.
\footnote{Triviality does not mean that field theories are useless for  descriptions of elementary particles and their interactions; they may not be free at all if a  finite (albeit large) ultraviolet cutoff is introduced. The lattice achieves precisely this and if it lies well beyond what is experimentally accessible  the theory is perfectly valid for as an accurate and mathematically well-defined model of elementary particle interactions, albeit at sufficient  energy.  Other theories such as pure QED and the standard SU(2) Higgs model  are also expected to be trivial~\cite{LW1}.  Nonetheless, triviality represents a defect in the theory as the cutoff has to be put in from the outside and cannot be fundamental. The Fields medal was awarded to Hugo Duminil-Copin in 2022 for proving mean-field critical behavior of the 4D Ising model and triviality of the 4D Euclidean scalar quantum field theory.}
    Martin L\"{u}scher and Peter Weisz were amongst those to the fore using the RG on this and, in a series of papers, addressed scaling laws and triviality bounds in lattice $\phi^4$ theory~\cite{LW1,LW2,LW3}. 
    Their review drew on earlier perturbation-theory results which were ``perhaps not as well-known to the lattice gauge community as they deserve to be''~\cite{LaKh69,BLZ}.
From these results, it was known that  $\hat{\nu}=1/6$ and $\hat{\gamma}=1/3$  so  $\hat{\qq} = 0$ would force $\hat\Delta = -1/2$ and standard FSS would force
$
 h_1 \sim L^{-\frac{\Delta}{\nu}}(\ln{L})^{\hat{\Delta}+\Delta\frac{\hat{\nu}}{\nu}}.
$  \index{finite-size scaling}
For the Ising universality class this would give $h_1 \sim L^{-3}(\ln{L})^{0}$.
The Q-alternative that $\hat{\qq}=1/4$ leads to $\hat{\Delta}=0$ so that $h_1 \sim L^{-3}(\ln{L})^{-1/4}$.
A negative exponent for the log term, compatible with this latter prediction,  
 was confirmed in Ref.~[\refcite{KeLa93}] using simulations of the 4D Ising model with $L=8$ to $24$ using the Swendsen—Wang cluster
algorithm.

Thus we conclude that scaling relations for logarithmic corrections from a modified Widom hypothesis adhering to the non-superlinear dogma are not correct. Instead, superlinearity as identified forty years ago by Br{\'e}zin is supported by numerics. 
Of course Br{\'e}zin's calculations and the numerics 
contained in Ref.~[\refcite{KeLa93}]  are for the Ising model in four dimensions with PBCs only. 
We will broaden these to far more general circumstances in what follows.

\subsection{Fisher zeros}

\index{zeros!Fisher}

A similar approach was used in Ref.~[\refcite{KeLa91}].
Written in terms of the Fisher zeros, the canonical partition function is 
\begin{equation}
Z_L(\beta) = \sum_{M=-dL^d}^{dL^d}   \rho_L(E) e^{\beta E}  ,
\label{gndcanIsing1t}
\end{equation} 
where the  spectral  density  $\rho_L(E)$ counts configurations with  given  values of $E$ only. 
The partition function is a polynomial in $y=\exp{(-\beta)}$ and the 
counterpart of Eq.(\ref{e153}) is 
\be
Z_L(\beta)=A_L(\beta)\prod_{j=1}^{dL^d}(y-y_j).
\label{e153t}
\ee

The free energy may then be written as a sum over Fisher zeros so that
\be
 f_{L}(\tau)= A^\prime_L(\tau) L^{-d}\sum_j\ln(\tau-\tau_j),
\ee
having again used conformal invariance to switch to the complex $\tau$ plane where the zeros are labelled $\tau_j$ and implicitly depend on $L$.
Differentiating appropriately wrt temperature, one may write the specific heat in terms of Fisher zeros as
\be
 c_L(t) = -L^{-d}\sum_j{\frac{1}{(\tau - \tau_j)^2}}.
 \label{Keith}
\ee
Again setting $\tau=0$ and taking the leading scaling to come from the first zero, one arrives at 
\be
 c_L(0) \sim L^{-d} \tau_1^{-2},
 \label{Zoe}
\ee
where $\tau_1$ is a measure of the difference between the first zero and the critical point.

 \index{finite-size scaling}
QFSS in the form (\ref{FSSpre}) applied to $c$ gives
\be
c_L(0) 
\sim 
L^{\frac{\qqqq \alpha}{\nu}}
(\ln{L})^{\hat{\alpha}+ \frac{\alpha(\hat{\qqqq}-\hat{\nu})}{\nu}}.
\label{11inPRL2}
\ee
To determine QFSS of the Fisher zeros we again appeal to the scaling ratio $x=\xi_L(0)/\xi_\infty (\tau)$ in Eq.(\ref{Q13}).
Expressing the partition function in terms of $x$, one finds $Z(\tau)=0$ when $\tau=\tau_1$ with
\be
\tau_1
\sim
L^{-\frac{\qqqq}{\nu}}
(\ln{L})^{ \frac{\hat{\nu}-\hat{\qqqq}}{\nu}}.
\label{click}
\ee
Combining Eqs.(\ref{Zoe}) and (\ref{click}) through Eq.(\ref{Zoe}), dimensionality $d$ again drops out of the leading scaling relations to recover $\nu d_{\rm uc} = 2 - \alpha$, and one recovers the scaling relation (\ref{Q11}): $\hat{\alpha} = d (\hat{\q}- \hat{\nu})$.\footnote{This is not the full story, however, and a subtlety arises in circumstances where the leading exponent $\alpha$ vanishes and the Fisher zeros inmpact at an angle of $\pi/4$. This happens in the 2D Ising model, for example. We refer the reader to Ref.~[\refcite{KennaPRL2} for details.}]

Again, ignoring $\q$ would give the standard hyperscaling form $\nu d = 2 - \alpha$, which fails above the upper critical dimension.

Ignoring  $\hat{\q}$ or setting $\hat{\q}=0$ in Eq.(\ref{click}), would deliver $ L^2\tau_1(L) \sim (\ln{L})^{1/3}$ for Eq.(\ref{Zoe}) when mean-field values of $\alpha$ and $\nu$ are inserted.
Incorporating $\hat{\q}=1/4$, on the other hand, delivers $L^2\tau_1(L) \sim (\ln{L})^{-1/6}$.
Indeed a negative logarithmic exponent compatible with this was observed for the 4D Ising model in Ref.~[\refcite{KeLa91}].
This establishes a positive (non-zero) value of $\hat{\qq}$ 

The inverse Eq.(\ref{Zoe}) above was used in Ref.~[\refcite{KeLa91}] to express the lowest-lying Fisher zeros  in terms of specific heat: 
\be
 \tau_1 \sim \frac{1}{\sqrt{L^d c_v}}.
\ee
There, this was used as an alternative approach to finding the FSS of Fisher zeros --- an approach not reliant on the QFSS form (\ref{Q13}). 
With $\alpha=0$, and the logarithmic exponent $\hat{\alpha}=1/3$ having again been provided by L\"{u}scher, Weisz and others~\cite{LW1,LW2,LW3,BLZ}, 
this also predicts that the first zero scales as $\tau_1 \sim L^{-2}(\ln{L})^{-1/6}$, as verified~\cite{Kennalog}.

A similar but more sophisticated approach was recently used by Aydin Deger and Christian Flindt~\cite{Deger1,Deger2,Deger3,Deger4}.
They used fluctuations of the total energy and the magnetisation to extract partition function zeros in systems with surprisingly small lattices (see also Ref.~\cite{moueddene2024critical} where this becomes a strategy to avoid resort to large-scale simulations which cost a lot in terms of carbon footprint).
The extra degree of sophistication provided by Deger and Flindt was to combine cumulants of different orders (not just the second-order cumulants used above).
Moreover, with their approach, ``critical exponents can be determined even if the system is away from the phase transition, for example at a high temperature.''
Finite-size scaling in high dimensions is addressed in Ref.~[\refcite{Deger3}].
The methods introduced by Deger and Flindt are powerful ways to extract partition function zeros from finite-size systems solely using fluctuations of thermodynamic observables and do not require prior knowledge of the partition function itself. 
Thus their approach opens new ways to access zeros numerically and in experiments.

Indeed, partition function zeros have found relevance for experiments only recently.
Long thought to be mathematical constructs without physical realization, in 2015 manifestations of imaginary fields were observed in magnetic resonance experiments performed on the spins of a molecule in Ref.~[\refcite{PhysRevLett.114.010601}]. 
The experiments followed a  theoretical proposal made by 
Bo-Bo Wei and Ren-Bao Liu a few years earlier~\cite{PhysRevLett.109.185701}.
They had considered an Ising spin bath and found an equivalence between its partition function with complex field and the quantum coherence of a probe spin coupled to the spin bath; ``the times at which the quantum coherence reaches zero are equivalent to the complex fields at which the partition function vanishes—that is, the fields that produce the Lee-Yang zeros''.
This was nicely verified in experiments by  Xinhua Peng and her colleagues Ref.~[\refcite{PhysRevLett.114.010601}].
See Ref.~[\refcite{Nerses}] for a non-technical summary of these important theoretical and experimental results.
\cite{PhysRevLett.114.010601}
\cite{PhysRevLett.109.185701}

\subsection{Pseudocriticality, shifting and rounding}

\index{pseudocritical point}\index{exponent!shift}\index{exponent!rounding}
In the discussion around Eqs.(\ref{eq-158}) and (\ref{sleep}), we purposefully relaxed notation.
In particular, we did not pay heed to any difference between the real and imaginary parts of the partition function zeros.
If our focus is on Lee-Yang zeros, in situations where the Lee-Yang theorem holds, the zeros necessarily lie on a singular line which is on the imaginary axis. 
I.e., the real part of the Lee-Yang zeros is zero for any lattice size. 
In this case, there is no ambiguity and $h_1(\tau)$ refers to the imaginary part of the first Lee-Yang zero.

In the case of Fisher zeros, however, the situation is slightly more complicated. 
Here $\tau_1$ has both real and imaginary parts which depend upon the system size $L$. 
Let us suppose the leading FSS behaviour for the real part is  ${\rm{Re}} [\tau_1(L)] \sim |\tau|^{-1/\nu_{\rm{real}}}$ and that for the imaginary part is as ${\rm{Im}} [\tau_1(L)] \sim |\tau|^{-1/\nu_{\rm{imag}}}$.
The partition function zero then scales as the slower of these two so that $\nu = {\rm{min}}(\nu_{\rm{real}},\nu_{\rm{imag}})$. 
The Fisher zeros impact onto the real axis at an angle given by 
$ \tan \phi \approx {{\rm{Im}} [\tau_1(L)]}/{{\rm{Re}}[ \tau_1(L)]}
  \sim  L^{\nu_{\rm{real}}-\nu_{\rm{imag}}}$.
  If $\nu_{\rm{real}}<\nu_{\rm{imag}}$, the impact angle would be zero in the infinite-volume limit, meaning a spread of singularities instead of a single critical point.
  If $\nu_{\rm{real}}>\nu_{\rm{imag}}$, on the other hand, the zeros impact vertically. 
  This means there is a symmetry between the ordered and disordered phases and the specific heat amplitudes on either side of the critical point have to coincide.
  Indeed, this happens in the 2D Ising model because of its self-dual property. 
  If $\nu_{\rm{real}}=\nu_{\rm{imag}}$, any angle of impact is possible. 
  For a full discussion of these matters, see Refs.~\refcite{GoKe11,Kennalog}].
  
  Therefore, in most circumstances the real part of the first Fisher zero scales to the infinite-volume critical point in the same way as the imaginary part approaches zero. The only circumstance where that might not be expected to happen is when the specific-heat amplitude ratio (a universal quantity) is one. 
  
  Moreover, Eq.(\ref{Zoe}) suggests that specific heat peaks when $\tau_1$ is smallest, i.e., when the temperature is close to the real part of the first Fisher zero.
  This means that the real part of the first zero is a pseudocritical point and this is shifted away from the critical point by an amount proportional to $L^{-\nu_{\rm{real}}}$, to leading order. 
  The location of the specific heat peak and the susceptibility peak are other pseudocritical points. 
  To encapsulate the more general nature of pseudocritical points, we follow traditional notation for the so-called shift\index{exponent!shift} exponent and denote it by $\lambda$
  (not to be confused with the generic critical exponent $\lambda$ used in the quotation by Fisher in Section~\ref{SecNotations}).

For a system of linear extent $L$, then, the pseudocritical point $\tau_L=T_L/T_c-1$ scales to leading order potentially with logarithms as
\begin{equation}
 |\tau|_L  \sim L^{\lambda} (\ln{L})^{\hat{\lambda}} ,
 \label{shift}
\end{equation}
where
\begin{equation}
 \lambda = \frac{1}{\nu}
 \label{shiftexp}
\end{equation}
provided the specific heat amplitude ratio is not one.

As stated, we do not wish to revisit the old ground covered in Volume III and Volume IV of this series. 
Instead, we simply summarise how these critical exponents are related\index{scaling relations} and refer the reader to Refs.~[\refcite{Kennalog,doi:10.1142/9789814632683_0001}] for details.
Suffice to say that the scaling relation associated with the logarithmic counterpart to the shift exponent $\lambda$ is
\begin{equation}
 \hat{\lambda}
 =
 -\hat{y}_t
 =
 \frac{\hat{\nu}-\hat{\qq}}{\nu}
.
\label{SRlogY}
\end{equation}
.

Pseudocriticality in this context can be defined in a number of ways and may even depend on the quantity $Q$ itself. 
Strictly speaking, we should use a notation such as  $\tau^Q_L$ to enforce this point. 
In that case, $\tau^\chi_L$ would vanish at the value of $T$ where the magnetic susceptibility reaches its peak value.
Likewise, $\tau^\xi_L$ would vanish at the value of $T$
where the correlation length peaks.
The correlation function doesn't have a peak, of course, and one way to identify it may be as the value of $T$ at which it becomes pure power law, possibly with logarithmic corrections.
Again, if we are to be strict, we should also introduce different notations for different reduced temperatures for different functions. Thus $\tau^{\chi}$ would represent $T/T^\chi_L-1$ and so on.
Thus there is no ambiguity in the definition of pseudocritical points or the reduced temperatures for finite-size systems.
We find the notation $\tau^Q$ and $\tau^Q_L$  with different $Q$'s unwieldy, however, so we use $\tau$ and $\tau_L$ throughout. The precise definition should be clear from the context.
Regardless of the definition of $\tau$ and $T_L$, they always revert to $\tau$ and $T_c$ in the thermodynamic limit.

Finally, we also mention the rounding exponent $\theta$ and its logarithmic counterpart $\hat{\theta}$.
This is associated with the smoothening out of an infinite-volume divergence into a finite-volume peak. 
One (rather arbitrary) measurement of rounding is the width of a particular curve at half of its maximum height. 
Thus the rounding associated with the susceptibility, for example, is the width of the susceptibility curve when susceptibility attains half its maximum value for finite size. 
We write
\begin{equation}
 \Delta T \sim L^{-\theta} (\ln{L})^{\hat{\theta}}
 \label{round}
\end{equation}
to leading order in logarithms.

\subsection{Summary}

At this stage of our considerations, we still have not paid attention to the details of any given underlying Hamiltonian. 
We have not even considered symmetries that lead to the Lee-Yang theorem or Fisher circles.
We have limited our presentation to circumstances where the partition function zeros fall on a singular line but even that is not a requirement for our considerations.
See Ref.~[\refcite{JJK}] for circumstances where the zeros are not restricted to a curve in the complex plane and/or come in degenerate sets.  

To take this further, we wish next to get to the heart of the reasons for superlinear correlation length and the origins of the $\qq$ and $\hat{\qq}$ exponents.
The issues we wish to address are, then,  the correlation sector,  hyperscaling, and finite-size scaling (FSS) in very general renormalization-group terms.
Our contribution to Volume III sets the framework: 
if $\hat{\qq}$ is required for self-consistency of FSS with logarithmic corrections, including at the critical dimension, what is the source and role of its counterpart \coppa~in higher dimensions?
This is where RG comes into play.
Until the work presented in Volume IV of this book series, dangerous irrelevant variables (DIV) had been considered only to apply to the free energy of a given model and related thermodynamic functions, which otherwise might fail to match the valid predictions of mean-field theory.
DIVs were believed not to manifest in the correlation sector which appeared not to fail.
The other two mechanisms, hyperscaling and FSS, were considered to fail in high dimensions.
Each of these three beliefs --- the non-failure of the correlation sector and the failure of both FSS and hyperscaling --- were overturned in Volume IV, in a move that endowed the renormalization group with a more complete degree of encompassment~\cite{doi:10.1142/9789814632683_0001}.
Here we extend those considerations to  $\phi^n$ theories and update them with the latest relevant literature.

Our purpose in this section is not to re-derive the full details of scaling relations for log corrections.
We refer to Refs.~[\refcite{Kennalog,HattedScalingLaws}] for a more comprehensive treatment.
Instead, our intent here is to give a pedagogic and self-contained account of superlinearity and the validity of QFSS - a framework around which a self-consistent theory must be built. 
In the above considerations we have seen how critical behaviour is driven by the first few Lee-Yang or Fisher zero. 
The gross approximations of dropping the summations in Eqs.(\ref{eq-158}) and (\ref{Zoe}) are justified posteriori --- it works.
For a more extensive exposition, higher-order zeros and the density of zeros along the singular line should be accounted for. 
We refer the reader to Refs.~[\refcite{KennaPRL1,KennaPRL2}] for such an account.

\section{Ginzburg-Landau theory (mean-field)}
\label{LMFT}

\index{Ginzburg-Landau!theory}\index{mean-field theory}
In Ginzburg-Landau theory, we consider physical systems  described in  thermal equilibrium  by the partition function
\be
 Z=\int {D}\phi \ \!e^{-F[\phi]},\label{eq-Z31}
\ee
where the functional 
\be
F[\phi]=\int d^dx\ \! f(\phi,\bnabla\phi) \label{Eq-32}
\ee
is a free energy (or action) integrated over all space and
\be
f(\phi,\bnabla\phi)={\textstyle \frac 12} r\phi^2(\vac x)+
{\textstyle \frac 13}u_3\phi^3(\vac x)
+{\textstyle \frac 14}u_4\phi^4(\vac x)
+{\textstyle \frac 16}u_6\phi^6(\vac x)
-h\phi(\vac x)+{\textstyle \frac 12}|\bnabla\phi|^2.
\label{Eq-33}
\ee
is a free energy density.
In presenting the free energy density as a power expansion of the order parameter and its derivatives, this expression captures the essence of a multitude of models in statistical physics and, as such, can be applied to many different systems.
In reference to any specific system, such as the Ising model itself, the Ginzburg-Landau-Wilson theory may be considered phenomenological.
According to the Oxford Learners' Dictionary, the word ``phenomenological'' pertains to ``the branch of philosophy that deals with what you see, hear, feel, etc. in contrast to what may actually be real or true about the world.''
In this sense, it is not fundamental. 
However, describing  $\phi^4$ theory in this way misses the point.
The  $\phi^4$ theory removes details such as the quantum nature of the spins behind the Ising model or the precise details of interaction strengths.
It strips these back to the bare minimum in terms of dimension and symmetry group that is required for a phase transition in the same universality class.
In the case of the Ising model, there is even an (almost)  exact mathematical transformation  (called the  Hubbard-Stratonovich transformation, see e.g. Ref.~\cite{Kopietz2010}) that maps the Ising model onto the corresponding field theory (``almost'' since a $\ln{\cosh}$ term is expanded to quartic order). 
Thus, in moving to the Ginzburg-Landau-Wilson model we are not losing any fundamental aspect of the theory - it is not a mere simplification of a specific model, it captures the real and true essence of what is dimensionality, symmetry and boundary conditions without worrying about non-universal details.

To avoid unnecessary clutter we present this chapter in terms of a scalar matter field with $O(1)$ symmetry rather than in terms of vector fields. The extension of the considerations presented in this chapter to the general $N$ case is similarly obvious.
These $O(N)$ theories involve higher values of $N$ and we refer the reader to 
Ref.~[\refcite{Ke2004}] for a finite-size scaling theory {\emph{at}} the upper critical dimension.

The coefficient $r$ in Eq.(\ref{Eq-33}) is a reparameterisation of the reduced temperature $\tau$ we had earlier. We take it as positive in the disordered phase and negative in the ordered one. This is required by consistency to have a vanishing order-parameter in the disordered phase as we will see below. 
In the case of percolation, for example, there is no temperature and $r$ is anyway preferred (over $\tau$).
We denote by  $u_n$ the coefficient of the highest power of $\phi$ in the above Lagrangian.
The case $n=3$ refers to percolation and the case $n=4$ with $u_3=0$ is the standard Ginzburg-Landau-Wilson model. 
As stated, it maps (almost exactly) to the Ising model.
The case $n=6$ is of a tricritical point which marks the singular behaviour at the end of a line of first-order phase transitions.
An example of this is found in the  Blume-Capel model~\cite{Blume,CAPEL,PhysRevA.4.1071,moueddene2024critical}.
Thus, despite trimming it back to bare essentials, the $\phi^n$ model indeed covers a range of universality classes.

A simplified version called the Ginzburg-Landau theory
 does not take into account fluctuations of the order parameter close to the critical point and in this sense, it is a mean-field theory where only averages count
 (an even simpler version in which the order parameter is spatially uniform is called the Landau theory).
Nonetheless, the theory is accurate in most of its predictions when the system is connected enough and this is the case for high-dimensional systems.
To access it, we identify the field configurations $\phi_0(\vac x)$ which have the highest weight,
\be
\left.\frac{\delta F}{\delta\phi}\right|_{\phi_0(\vac x)}=0
\label{eq-33}
\ee
from which
\be
\frac{\partial f}{\partial\phi}-\bnabla\cdot\frac{\partial f}{\partial (\bnabla\phi)}=0,\quad\hbox{at}\quad \phi=\phi_0(\vac x).
\label{ELE}
\ee
The gradient term in Eq.(\ref{Eq-33}) does not have an associated coefficient because it has been absorbed into the other coefficients to render the term dimensionless once integrated over space. 
 Therefore, the matter-field scaling dimension,  $x_\phi$, is
\be 
x_\phi =\frac d2-1.
\label{eq-36a}
\ee
Compare this to Eq.(\ref{defEta}), which was derived from power counting.
We see that the $\eta$ exponent vanishes for mean-field theory.

We consider the $\phi^n$, then, in its generic form: 
\be
f(\phi,\bnabla\phi)={\textstyle \frac 12} r\phi^2(\vac x)+
{\textstyle \frac 1n}u_n\phi^n(\vac x)
-h\phi(\vac x)+{\textstyle \frac 12}|\bnabla\phi|^2.
\label{eq-36}
\ee
Dropping the gradient term for an  infinite homogeneous system, Eq.(\ref{eq-33}) gives 
\be
\phi_0(r+u_n\phi_0^{n-2})=h.
\label{eq-37}
\ee
If $h=0$ we identify the order parameter $m_\infty$ as  $\phi_0$ so that
\bea
m_\infty(  \tau)=({-r}/{u_n})^\frac{1}{n-2}, &\quad& T<T_c,   \label{eq-38}\\
m_\infty(  \tau)=0,                          &\quad& T>T_c.   \label{eq-39}
\eea
From this we extract the critical exponent 
\be
\beta_{\rm\scriptscriptstyle MFT}=\frac 1{n-2}.
\label{eq-41} 
\ee
and magnetisation amplitude in the ordered phase, as used in Eq.(\ref{defB}), as then $B^-=(u_n)^{-1/(n-2)}$. \index{critical!amplitude}
The magnetic field dependency of the order parameter at the critical temperature $r=0$ also comes from  Eq.(\ref{eq-37}) and is
\bea
m_\infty(h)=\hbox{sgn}\ \!(h)\ \!({|h|}/{u_n})^{\frac{1}{n-1}}, 
&\quad&
T=T_c,
\label{eq-42}
\eea
from which, comparing to Eq.(\ref{defBc}),
\be
\delta_{\rm\scriptscriptstyle MFT}=n-1\label{eq-43}
\ee
together with the amplitude $D_c=(u_n)^{-1/(n-1)}$.

The second derivative of Eq.(\ref{eq-37}) wrt $h $ gives the  susceptibility 
\be\chi_\infty=[r+(n-1)u_n\phi_0^{n-2}]^{-1}.\ee
Eqs.(\ref{eq-38}) and (\ref{eq-39}) then give
\bea
\chi_\infty(  \tau)=[{(n-2)(-r)}]^{-1},  &\quad&  T<T_c, \\
\chi_\infty(  \tau)= r^{-1},  &\quad&  T>T_c, 
\eea
for the two phases.
Both of these deliver the same critical exponent [see Eq.(\ref{defGamma})]
\be
\gamma_{\rm\scriptscriptstyle MFT}=1,
\label{eq-47}
\ee
and the associated amplitudes are $\Gamma^-=
(n-2)^{-1}$ and $\Gamma^+=1$.

The free energy is given by inserting the equilibrium order parameter  (\ref{eq-38}) and (\ref{eq-39}) in the expansion (\ref{eq-36}), so that
\bea
 f_\infty(  \tau)=\left(\frac 1n-\frac 12\right)u_n^{\frac{2}{2-n}}(-r)^{\frac n{n-2}},  &\quad& T<T_c,
 \label{eq-47bis}\\
 f_\infty(  \tau)=0, &\quad&  T>T_c.
\eea
The second temperature derivative gives the 
 specific heat exponent from equation (\ref{defA}) as
\bea
 c_\infty(  \tau)=\frac 1{2-n}u_n^{\frac{2}{2-n}}(-r)^{\frac {4-n}{n-2}},  &\quad& T<T_c, \label{eq-47ter}\\
 c_\infty(  \tau)=0, &\quad& T>T_c.
\eea
Thus a jump appears at the phase transition and the exponent is associated with the low-temperature regime only (for this, the mean-field solution).
Notwithstanding this, we write,
\be
\alpha_{\rm\scriptscriptstyle MFT}=\frac{n-4}{n-2},\label{eq-52}
\ee
with the amplitude in this regime given by 
\be (A^-/\alpha_{\rm\scriptscriptstyle MFT})=(u_n)^{-2/(n-2)}/(2-n).\ee

For the correlations, one has to reinstate the gradient term to Eq.(\ref{eq-36}). 
The Euler-Lagrange equation (\ref{ELE}) then leads to 
\be
r\phi(\vac x)-u_n\phi^{n-1}(\vac x)-\bnabla^2\phi(\vac x)=h.\label{eq-52bis}
\ee
The space dependency of the correlation function can be extracted from the order parameter profile when a localized magnetic field
$h_0\delta(\vac x)$  is applied at the origin.
At criticality $r=h=0$,  and $\phi(\vac x)$ can be considered small enough to neglect the non-linear term. Outside the origin, this leads to a Laplace equation which happens to be independent of the value of $n$ at this level of approximation.
One finds (assuming isotropy)
\be
\bnabla^2\phi(\vac x)= \frac{1}{|\vac x|^{d-1}} \frac d{d|\vac x|} \left(|\vac x|^{d-1}\frac{d\phi(\vac x)}{d|\vac x|} \right)= 0.\ee
The solution takes the form
\be
g(\vac x)\sim\frac{1}{|\vac x|^{d-2}}.
\ee
This is consistent with the critical exponent
\be
\eta_{\rm\scriptscriptstyle MFT}=0.\label{eq-56}
\ee
Again, this does not depend on $n$.
Beyond the critical temperature, [say above $T_c$, since we are still neglecting the $\phi^{n-1}$ term in Eq.(\ref{eq-52bis})]\footnote{Below the critical temperature we may assume the same $\tau$ dependency for the correlation length  (an argument that can be made more rigorous).~\cite{cha95}
}
 the equation that has to be solved is
\be
r\phi(\vac x)-\bnabla^2\phi(\vac x)=h.\label{eq-56bis}
\ee
The only one at our disposal in this, the infinite-volume, thermodynamic, limit is the correlation length.
For this reason, we identity the correlation lengths $\xi(\tau,h=0)$ and $\xi(\tau=0,h)$ 
\be
\xi(\tau,h=0) \sim 1/|r|^{1/2},\qquad 
\xi(\tau=0,h)\sim |\phi/h|^{1/2},
\label{eq-xiapprox}
\ee
for the two different phases.
 With $\phi \sim h^{1/\delta}$, both  exponents of the correlation length follow:
\be
\nu_{\rm\scriptscriptstyle MFT}=1/2,\qquad
\nu_{\rm\scriptscriptstyle c\ \!MFT}=
\frac{\delta - 1}{2\delta} = \frac{n-2}{2(n-1)}.\label{eq-59}
\ee

For convenience, we collect the mean-field exponents for the Ising, percolation and tricriticality universality classes in Table~\ref{tab2}. \index{Gaussian!model}\index{Gaussian!fixed point}

\begin{table}[ht]
\tbl{Critical exponents for the Gaussian model and mean-field critical exponents for percolation,  Ising magnets (the SAW has the same exponents) and for tricriticality as well as for the generic Landau model.
RG eigenvalues are $y_t = 2$, $y_h=d/2+1$ and $y_u=n+(2-n)d/2$.
The latter is negative when $d > d_{\rm uc}$.
Despite this irrelevancy of the field $u$, only $\gamma$, $\nu$ and $\eta$ match the Gaussian prediction.
}
{\begin{tabular}{@{}llcccccccc@{}} \toprule
Model          & $\phi^n$ 
                            & $\alpha$ 
                                & $\beta$ 
                                    & $\delta$ 
                                        & ${\nu_c}$ 
                                            & $\gamma$ 
                                                &  $\nu$ 
                                                    & $\eta$ 
                                                      & $d_{\rm uc}$ \\            
\hline
GFP            & $\phi^2$ 
                            &  $2-\frac{d}{2}$
                                &  $\frac{d-2}{4}$
                                    &  $\frac{d+2}{d-2}$   
                                        &  $\frac{2}{d+2}$  
                                            &  $1$
                                                &  ${\textstyle \frac 12}\vphantom{{\displaystyle \frac 12}}$              
                                                   & $0$ 
                                                      &  $\frac{2n}{n-2}$
                                                  \\ 
Percolation    & $\phi^3$ 
                        & $-1$          
                            & $1$ 
                                & $2$ 
                                    & $\frac{1}{4}$  
                                        & $1$ 
                                            & $\frac{1}{2}$ 
                                                & $0$ 
                                                    & $6$  \\
Magnets, SAW   & $\phi^4$ 
                            & $0$  
                                & ${\textstyle \frac 12}$ 
                                    & $3$  
                                        & $\frac{1}{3}$   
                                            & $1$ 
                                                & ${\textstyle \frac 12}\vphantom{{\displaystyle \frac 12}}$ 
                                                    & $0$  
                                                       & $4$\\
Tricriticality & $\phi^6$ 
                            & ${\textstyle \frac 12}$ 
                                & ${\textstyle \frac 14}$ 
                                    & $5$  
                                        & $\frac{2}{5}$ 
                                            & $1$ 
                                                & ${\textstyle \frac 12}\vphantom{{\displaystyle \frac 12}}$ 
                                                   & $0$  
                                                      & $3$ \\

Landau         & $\phi^n$ 
                            & $\frac{n-4}{n-2}$ 
                                & $\frac{1}{n-2}$  
                                    &$n-1$
                                        & $\frac{n-2}{2(n-1)}$ 
                                            & $1$ 
                                                & ${\textstyle \frac 12}\vphantom{{\displaystyle \frac 12}}$ 
                                                    & $0$  
                                                      & $\frac{2n}{n-2}$ \\                                                  
\botrule                                                      
\end{tabular}}
\label{tab2}
\end{table}

Finally, we reflect on where the above considerations are valid.
Fluctuations are measured by the susceptibility, of course, and this is an integral over space of the correlation function:
\be
\chi\simeq\int d^dx\ \! g(\vac x)\sim |  \tau|^{-\gamma_{\rm\scriptscriptstyle MFT}}.
\ee
The square of the magnetisation  inside the correlation volume, on the other hand, is 
\be
\xi^d m_\infty^2\sim|  \tau|^{-d\nu_{\rm\scriptscriptstyle MFT}+2\beta_{\rm\scriptscriptstyle MFT}}.
\ee
These two quantities have  the same dimensions and, comparing them, fluctuations are relatively weak if  $d \ge d_{\rm uc}$ where
\be
d_{\rm uc}= \frac{2\beta_{\rm\scriptscriptstyle MFT}+\gamma_{\rm\scriptscriptstyle MFT}}{\nu_{\rm\scriptscriptstyle MFT}}.
\ee
The inequality $d \ge d_{\rm uc}$ is the Ginzburg criterion --- a neat indicator of where order parameter fluctuations can be neglected and where mean-field theory kicks in. 
Inserting Eqs.(\ref{eq-41}), (\ref{eq-47}) and (\ref{eq-59}) for the mean-field exponents, we then get
\be 
d_{\rm uc}= \frac{2n}{n-2}.
\label{ginzy}
\ee
The demarcation point  $d_{\rm uc}$ is the upper critical dimension and its values for different models are given in Table~\ref{tab2}.

\section{The Gaussian fixed point: its apparent sufficiency for the correlation sector and its insufficiency for the free-energy sector}
\label{Gscaling}

\index{Gaussian!fixed point}
The dimensionlessness of the partition function in Eq.(\ref{eq-Z31}) dictates the 
dimensionality of each individual term in  the free energy density 
(\ref{Eq-33}) or (\ref{eq-36}).
From this we obtain the scaling dimension of  the matter field as Eq.(\ref{eq-36a}) and
\bea
&y_  t+2x_\phi = d, &  {y_t}=2, \label{eq-62}\\
&{y_h}+x_\phi = d, &  {y_h}=\frac d2+1,\label{eq-63}\\
&y_{u}+nx_\phi = d, &  y_{u}=\frac d2(2-n)+n 
.\label{eq-64}
\eea
Here, and henceforth, we use $u$ for $u_n$ and $y_u$ for $y_{u_n}$).
The renormalization flow  $\tau'=b^{{y_t}}\tau$, $h'=b^{{y_h}}h$, $u'=b^{y_{u}}u$ is controlled by these  RG  eigenvalues, and
 $\tau=h=u=0$ is identified as the Gaussian Fixed Point (GFP).
 This is because there the partition function (\ref{eq-Z31}) becomes a Gaussian integral,
\be
Z=\int D\phi\ \!e^{-\int d^dx\ \!\frac 12|\bnabla\phi|^2}.
\ee
The additional scaling field $u$ takes us beyond the simple  homogeneous form (\ref{eq-10}) to
\be
f_\infty^{\rm sing}(  \tau,h,u)=b^{-d}f_\infty^{\rm sing}( b^   {{y_t}}  \tau, b^   {y_h} h, b^   {y_u} u) .
\label{eq-10new}
\ee 
Likewise, the correlation function and correlation length take the forms
\be
g_\infty^{\rm sing}(\vac x,  \tau,h,u)=b^{-2x_\phi}g_\infty^{\rm sing}(b^{-1}\vac x , b^   {{y_t}}  \tau, b^   {y_h} h, b^   {y_u} u)\label{eq-10new2}
\ee 
and
\be
\xi_\infty^{\rm sing}(  \tau,h,u)=b \xi_\infty^{\rm sing}( b^   {{y_t}}  \tau, b^   {y_h} h, b^   {y_u} u),\label{eq-10new3}
\ee 
respectively.

The RG eigenvalues $y_t$ and $y_h$ in Eqs.(\ref{eq-62}) and (\ref{eq-62}) are positive so their associated scaling fields are relevant in the sense that they drive the system away from the fixed point ($\tau=h=u=0$) as the RG is applied. 
I.e., under rescaling by a factor $b>0$, they grow as $\tau \rightarrow \tau^\prime$ and $h \rightarrow h^\prime$ with
\be
  \tau^\prime=b^{2}  \tau,\quad\hbox{and}\quad h^\prime =b^{\frac d2+1}h.
\ee
While the scaling field $u$ can also have a positive RG eigenvalue in  Eq.(\ref{eq-64}), this is only the case when the Ginzburg criterion (\ref{ginzy}) fails.
However, when the Ginzburg criterion holds, $y_{u}<0$ and starting from any non-zero value of $u$ drives the system back to the critical point. 
For this reason, $u$  is said to be irrelevant above the upper critical dimension --- its inclusion does not affect the universality class (critical exponents) of the model. 
Precisely at $d_{\rm uc}$, one has the marginal situation where multiplicative logarithmic corrections to scaling arise. 

Therefore, naively (or without paying attention to fifty years of literature), 
one might expect the predictions from RG at the GFP to match those coming from Landau mean-field theory of the previous section and we also list them in Table~\ref{tab2} for ease of comparison.
We extract these by inserting the scaling dimension (\ref{eq-36a}) and the RG eigenvalues (\ref{eq-62}) and (\ref{eq-63}) into 
Eqs.(\ref{defBc})--(\ref{defEta}).

There are a number of interesting observations to make here.
Firstly the exponents $\gamma$, $\nu$, $\eta$ and the critical dimension all match the correct values coming from MFT.
Secondly, all of these values are independent of $n$.
This, of course, is because the scaling dimensions (\ref{eq-62}) and  (\ref{eq-63}) are independent of $n$ and, while the $n$-dependency lives only in $y_u$ [Eq.(\ref{eq-64})], that is irrelevant for the GFP above the upper critical dimension. 
The critical exponents listed for percolation, the Ising model and tricriticality have numerous verifications in the literature, confirming that the Landau exponents and not the GFP predictions are the correct ones.

All critical exponents listed in the table obey the scaling relations (\ref{alphabetagamma})--(\ref{oldeta}) as well as (\ref{extraone}), derived from simple homogeneous assumptions outlined in Sec.\ref{homo}. 
The hyperscaling relation (\ref{oldhyper}) only holds for the GFP. 
Although it holds precisely at the upper critical dimension for percolation, Ising magnets, tricriticality and the general Landau scheme, it does not hold for general $d>d_{\rm uc}$.
\index{critical!exponent}\index{exponent!critical}

The key to unravelling why the GFP fails to match Landau theory above $d_{\rm uc}$, and why hyperscaling fails in Landau theory there too,  is the observation that the wrong exponents $\alpha$, $\beta$ and $\delta$ all come from derivatives of the free energy, while the correct $\gamma$, $\nu$ and $\eta$ are each associated with correlations. 
This is why we distinguish between the  ``free energy sector'' and  ``correlation sector'' in the introduction to this chapter and the title of this section. 

There are two exceptions to the rule that are extremely informative.
Firstly, although the susceptibility belongs to the free energy sector, the GFP delivers the correct mean-field value of the critical exponent for it. 
This is because, besides being a direct derivative of the free energy, the susceptibility can also be extracted by integrating the correlation function. Therefore it also belongs to the correlation sector. 
This suggests the robustness of the correlation sector as delivering the correct Landau exponents from the GFP.
However, the GFP value for 
$\nu_{\rm\scriptscriptstyle c}$
is $\nu_{\rm\scriptscriptstyle c\ \! G} = 2/(d+2)$ from inserting Eq.(\ref{eq-63}) for $y_h$ into 
Eq.(\ref{defNub}). 
This does not agree with the mean-field value in Eq.(\ref{eq-59}) except when $d=d_{\rm uc}$ (where all exponents coincide). This suggests that something is also wrong in the correlation sector!
This observation, along with the search for the leading counterpart of $\hat{\qq}$ guides the way to open new insights into high dimensions that lay hidden since the inception of RG.

\section{Dangerous irrelevancy: Rescuing the free-energy sector in 
infinite volume and attempts at resuscitation in finite volume}
\label{DIVscene}

\index{dangerous irrelevant variable}
At this point, we have seen that, while the self-interacting field $u$ is irrelevant above the upper critical dimension, the GFP is only partially successful in its delivery of critical exponents there. Moreover, that success lies in the correlation sector only. Also, while hyperscaling matches the GFP, it fails for Landau theory when the Ginzburg criterion holds.
Next, we follow Fisher and seemingly rescue the situation for the thermodynamic limit before we encounter another obstacle in finite systems.

\subsection{Dangerous irrelevant variables for thermodynamic functions at infinite volume}
\label{breakthrough}

Gathering the seemingly innocuous amplitudes of Eqs.(\ref{eq-38}), (\ref{eq-42}) and (\ref{eq-47ter}), namely \index{critical!amplitude}
\bea
&&B^-=u^{-\frac 1{n-2}},\label{eq_76}\\
&&D_c=u^{-\frac 1{n-1}},\label{eq_77}\\
&&\frac{A^-}{\alpha_{\rm\scriptscriptstyle MFT}}=\frac1{2-n}u^{-\frac 2{n-2}},
\label{eq_78}
\eea
 we spot a danger: they are each singular when $u\to 0$!
Michael Fisher highlighted this in the 1980's~\cite{FisherStellenbosch} although he may have noticed it ten years earlier~\cite{Gunton1973RenormalizationGI}. 
Therefore, although ``irrelevant'' in the sense of RG flow, the field $u$ is  {\em dangerous} for the amplitudes. 
The amplitudes of the usual quantities in the correlation sector are not $u$-dependent and do not face the same danger. 
An exception to this observation is  $\xi_c$,  the $h-$dependent correlation length at $T_c$.
This quantity is seldom discussed, however, so, as many others have done, we leave it aside for the moment.
Suffice it to say for now that $u$ is a  {\em dangerous irrelevant variable} (DIV) for the free-energy sector.

If we define the exponents appearing in Eqs.(\ref{eq_76})--(\ref{eq_78}) as
\be
\kappa=\frac 1{n-2},\quad\lambda=\frac 1{n-1},\quad\mu=\frac 2{n-2},
\label{E81}
\ee
we may write the free-energy derivatives as
\bea
 &&m_\infty(  \tau<0,h=0,u)\sim  |\tau|^\beta u^{-\kappa}, \\
 &&m_\infty(  \tau=0,h,u)\sim |h|^{1/\delta} u^{-\lambda},\\
 && c_\infty(  \tau,h=0,u)\sim |  \tau|^{-\alpha}
u^{-\mu}.
\eea

Eqs.(\ref{eq-1a})-(\ref{eq-1e}) have clearly to be modified to take the DIVs into account and we write
\bea
&&m_\infty(  \tau,0,u)
\stackrel{u\to 0}{=}b^{-d+{y_h}-\kappa y_{u}}u^{-\kappa}{\mathscr M}^-(b^{y_t}\tau,0),\label{eq-81}\\
&&m_\infty(0,h,u)\stackrel{u\to 0}{=}b^{-d+{y_h}-\lambda y_{u}}u^{-\lambda}{\mathscr M}_c(0,b^{y_h}h),\\
&&c_\infty(  \tau,0,u)\stackrel{u\to 0}{=}b^{-d+2{y_t}-\mu y_{u}}u^{-\mu}{\mathscr C}^\pm(b^{y_t}\tau,0) \label{eq-85}.
\eea
Fixing $b$ to  $|  \tau|^{-1/   {y_t}}$ or 
$|h|^{-1/   {y_h}}$ in (\ref{eq-81})--(\ref{eq-85}), appropriately then gives~\cite{Goldenfeld}
\bea
\beta	_{\rm\scriptscriptstyle MFT}=\frac{d-   {y_h}}{   {y_t}}+\frac{\kappa y_u}{   {y_t}}
={\frac{1}{n-2}},
\\
\frac{1}{\delta_{\rm\scriptscriptstyle MFT}}
=\frac{d-   {y_h}}{   {y_h}}+\frac{\lambda y_u}{   {y_h}}
={\frac{1}{n-1}},
\\
\alpha_{\rm\scriptscriptstyle MFT}
=\frac{2   {y_t}-d}{   {y_t}}-\frac{\mu y_u}{   {y_t}}
={\frac{n-4}{n-2}}.
\label{E85}
\eea
These now match Eqs.(\ref{eq-41}), (\ref{eq-43}) and (\ref{eq-52}) so the free energy sector in the RG formalism above $d_{\rm uc}$ is repaired.
The amplitudes have also to be  consistent and we find, for example, $
B^-={\mathscr M}^-(0^-) u^{-\kappa}$.

Following the principle of not repairing that which does not appear to need repairing, nothing has to be modified for the correlation sector.

We have now reached a point where RG appears to be successful in its treatment of critical properties above the upper critical dimension in the thermodynamic limit {at least}. 

\subsection{A problem with Finite-Size Scaling}

There is, however a problem with FSS.  \index{finite-size scaling}
Inserting the MFT exponents in the FSS prescription~(\ref{eq87}), one expects at the pseudo-critical point $t=0$  
\bea
&&{c_L(t=0,0)\sim L^{\frac{\alpha_{\rm MFT}}{\nu_{\rm MFT}}} = L^{2\frac{n-4}{n-2}}},\\
&&\chi_L(t=0,0)\sim L^{\frac{\gamma_{\rm MFT}}{\nu_{\rm MFT}}} =L^2,\\
&&m_L(t=0,0)\sim L^{-\frac{\beta_{\rm MFT}}{\nu_{\rm MFT}}} =L^{-{\frac{2}{n-2}}},\\
&&\xi_L(t=0,0) \sim L. \label{Eq96}
\eea
We call this {\emph{Landau FSS}} because the exponents which appear in powers of $L$ are all ratios of exponents from Landau theory.
Here we have presented FSS at the pseudo-critical point $t=0$ (see Eq.(\ref{EQ_tvsT_L})) rather than the critical point $\tau = 0$ (see Eq.(\ref{EQ_tauvsT})).
This is usually easier to work with numerically because simulations, like pseudo-critical points, necessitate finite size.  
In contrast, the critical temperature itself requires extrapolation to the limit of infinite volume. 
Usually (for example below the critical dimension), it makes little difference whether we implement FSS at $t=0$ or $\tau=0$.
This is because both of them fall within the same scaling window - they exhibit the same FSS behaviour.
This is the case if the rounding is not too narrow. 

These standard FSS predictions, however, fail to match numerical simulations and exact results.
In the early 1980's, Br\'ezin~\cite{Brezin82} considered finite-size correlation length for the $\phi^4$ model above $d_{\rm uc}=4$ on a  hypercubic systems with PBCs. 
His theoretical analysis predicted 
\be 
\xi_L(\tau=0,0)\sim L^{d/4}
\label{EQ_95}
\ee
which contradicts Landau scaling (\ref{Eq96}). 
Then, in 1985,  with Jean Zinn-Justin,  
Br\'ezin presented a more complete analysis for the $\phi^4$ model~\cite{BREZIN1985867}, delivering
\be \chi_L(\tau=0,0)\sim L^{d/2},\label{EQ_97}\ee
for the susceptibility behaviour, above $d_{\rm uc}$.
In the same year, Binder presented numerical results for the susceptibility of the  Ising model in 5D with PBCs~\cite{Binder85} (see also \cite{Rickwardt}). 
At the pseudo-critical point he obtained
\be \chi_L(  t=0,0)\sim L^{5/2},\quad\hbox{and}\quad    |\tau_L|=|T_L/T_c-1|\sim L^{-5/2}.
\label{E95}
\ee
This is in agreement with the results of Br\'ezin and  Zinn-Justin, but disagrees with Landau FSS.

Thus, despite Fisher's DIVs rescuing the GFP RG formalism above $d_{\rm uc}$, we arrive at a disagreement between Landau FSS and exact/numerical results.
Even though we know that GFP values for the critical exponents do not match MFT, attempts to invoke them for standard FSS do not come to the rescue. 
While they deliver different FSS in the free-energy sector (namely, $c_\infty \sim L^{\alpha_G/\nu_G}=L^{4-d}$ and $m_\infty \sim L^{-\beta_G/\nu_G}= L^{-(d-2)/2}$), they necessarily deliver the same as Landau FSS for $\chi$ and $\xi$ because the correlation sector is not (yet) perceived to be in danger. 
As we have seen, these are independent of $n$ and do not agree with exact or numerical results.

Therefore, while RG may have been rescued above $d_{\rm uc}$, by 1985 FSS was to be sacrificed.
We next address an immediate attempt at resuscitation.

\subsection{Dangerous Irrelevant Variables for finite size:   the ``starred'' Renormalization Group eigenvalues}
\label{DIVFSS}

\index{dangerous irrelevant variable}
The consequences of the existence of DIVs for the finite-size scaling form of the free energy, and for hyperscaling, were investigated in 1985 in  Ref.~[\refcite{PhysRevB.31.1498}]. 
There, with Binder and Young, Michael Nauenberg and Vladimir Privman suggested to extend Fisher's ideas so that 
not only the observable thermodynamic functions but also the underlying free energy might be affected.
We refer to the authors of this important paper as BNPY.
In particular, BNPY suggested that
 {[cf Eq.(\ref{eq-10})]:}
 \be 
f_\infty(x,y,z)\stackrel{{z}\to 0}{=}z^{p_1}f_\infty(xz^{p_2},yz^{p_3},0)
\label{BNPYL1}
\ee
or
\be
f_\infty(  \tau,h,u)\stackrel{u\to 0}{=}b^{-d+p_1y_u}u^{p_1}{\mathscr F}^\pm( b^{   {y_t}+p_2y_u}  \tau u^{p_2}, b^{   {y_h}+p_3y_u} hu^{p_3}),
\label{eq-102}
\ee 
where $p_1$, $p_2$ and $p_3$ are constants feeding into $\kappa$, $\lambda$ and $\mu$ used above. 
BNPY also introduced a ``starred'' notation for the renormalised RG eigenvalues:
\be d^*=d-p_1y_u,\quad
y_  t^*=   {y_t}+p_2y_u,\quad y_h^*=   {y_h}+p_3y_u.
\label{E104}
\ee

For the $\phi^n$ model of equation (\ref{eq-36}), differentiation of the DIV-affected free energy density wrt $\tau$, and to $h$ allows
\bea
&\displaystyle \alpha_{\rm\scriptscriptstyle MFT}=\frac{2   {{y_t^*}}-d}{   {{y_t^*}}}, 
&\beta_{\rm\scriptscriptstyle MFT}=\frac{d-   {{y_h^*}}}{   {{y_t^*}}},
\label{SH1}\\
&\displaystyle \gamma_{\rm\scriptscriptstyle MFT}=\frac{2  {{y_h^*}} -d}{  {{y_t^*}}},
&\delta_{\rm\scriptscriptstyle MFT}=\frac{   {{y_h^*}}}{d-   {{y_h^*}}}.
\label{SH2}
\eea
Comparing with Fisher's  (\ref{E85}) in Subsection~\ref{breakthrough}, leads to 
\be
{p_1=0}, \quad  p_2=-\frac{2\kappa   {y_t}}{d+2\kappa y_u} {=-\frac{2}{n}}, \quad p_3=-\frac{\kappa(2   {y_h}-d)}{d+2\kappa y_u} {=-\frac{1}{n}},\label{E114}
\ee 
having used Eq.(\ref{E81}) for the general $n$ case.
Inserting these values in Eq.(\ref{E104}) gives
\be 
d^*=d,\quad
y_ t^*=   {y_t}-\frac{2y_u}{n},\quad y_h^*=   {y_h}-\frac{y_u}{n}.
\label{E1041}
\ee
In terms of $d$ and $n$ the starred scaling dimensions are
\be 
y_t^* = \frac{d(n-2)}{n}, \quad
y_h^* = \frac{d(n-1)}{n}, \label{E1041b}
\ee
and in terms of $\qq$ they read as
\be 
y_t^* =2\qq, \quad
y_h^* = \frac d2+\qq. \label{E1041bqoppa}
\ee
By construction, inserting $y_t^*$ for $y_t$ and $y_h^*$ for $y_h$ in (\ref{defBc}), (\ref{defB}), (\ref{defGamma}) 
and (\ref{defA}), delivers the correct Landau MFT critical exponents for $\alpha$, $\beta$, $\delta$ and $\gamma$ in (\ref{eq-41}), (\ref{eq-43}), and (\ref{eq-47}), (\ref{eq-52}).
Again by construction, the remaining main critical exponents $\nu$ and $\eta$ are left intact because the correlation sector is untouched in Eq.(\ref{eq-102}).
\index{critical!exponent!starred}\index{exponent!starred}

Having extended DIVs to the finite-size regime, DIVs also extend finite-size scaling for the susceptibility and magnetisation in terms of the starred exponents instead of the GFP ones.
In particular, they find that
\[
 m_L \sim L^{y_h^*-d}
\]
and
\[
 \chi_L \sim L^{2y_h^* - d}.
\]
These are indeed correct as we shall see.
However, in BNPY's formalism, they do not come from FSS in a form like  Eq.(\ref{FSS1}).

BNPY made an insightful step analogous to Eq.(\ref{eq-102}) for the correlation length.
They proposed  that
\be
\xi_\infty(  \tau,h,u)\stackrel{{u\to 0}}{=}b^{1+q_1y_u}u^{q_1}{\Xi}( b^{   {y_t}+q_2y_u}  \tau u^{q_2}, b^{   {y_h}+q_3y_u} h u^{q_3}),
\label{eq-104}
\ee 
with the three  parameters,  $q_1$, $q_2$ and $q_3$ initially left open.
Then they make the following crucial but erroneous assumption: 
\begin{quote}{ ``Since the finite-size correlation length $\xi_L$ is bounded by $L$, we require $q_1 y_u<0$. {\rm (}\dots{\rm )} if one adopts the plausible assumption that for $  \tau=h=0$, the correlation length increases up to the linear dimensions of the lattice, which implies that $q_1=0$.''}
\end{quote}
We will see later that the value $q_1=0$ is not correct.
It feeds into the exponent $\qq$ in an essential way --- the assumption that $q_1=0$ is equivalent to assuming $\qq=1$ as we shall see.
BNPY did not pursue a discussion of $q_2$ or $q_3$, leaving open the possibility that they might differ from $p_2$ and $p_3$. 
We shall see below that, although this option has to be considered, this is not the case. 

At this stage, we are still confronted with the unsatisfactory situation that, 
while all critical and finite-size scaling behaviour seems to be rescued, the FSS prescription itself fails above the upper critical dimension.
In an attempt to replace that, 
 a new length scale was introduced --- to replace the (seemingly unbroken), correlation length, 
as the relevant scale which controls finite-size effects there.
Several other length scales were invented by other authors and these are discussed in Volume~IV~\cite{doi:10.1142/9789814632683_0001}. 
We do not pursue these considerations here as we believe them to be redundant.
Instead, we follow the outstanding work by  Luijten and  Bl\"ote~\cite{PhysRevLett.76.1557,PhysRevLett.76.3662.3}, which uses RG to bring corrections to scaling into the DIV scenarios and to connect with much of the mathematical literature.

\section{Scaling of the Fourier modes }
\label{modes}

\index{Fourier modes}
Eq.(\ref{eq-36}) gives the free energy for the generic  $\phi^n$ model, and its $n=4$ realisation is 
\be
F_{\rm GLW}[\phi(\vac x)]=\int d^dx \Bigl({\textstyle \frac 12} r\phi^2(\vac x)+
{\textstyle \frac 14}u\phi^4(\vac x)
-h\phi(\vac x)+{\textstyle \frac 12}|\bnabla\phi|^2\Bigr).\label{E-159}
\ee
If PBCs are used, this can be expressed in Fourier space
as
\bea
F_{\rm GLW}[\tilde\phi_{\vac k}]&=&\half\sum_{\vac k}(r+|\vac k|^2)|\tilde\phi_{\vac k}|^2
-
hL^{d/2}\tilde\phi_0
\nnb\\ &&
+
\quarter uL^{-d}\sum_{\vac k_1,\vac k_2,\vac k_3}\tilde\phi_{\vac k_1}\tilde\phi_{\vac k_2}
\tilde\phi_{\vac k_3}
\tilde\phi_{-(\vac k_1+\vac k_2+\vac k_3)} \,  \label{E161}
\eea
with 
\be
\phi(\vac x)=\sum_{\vac k}\tilde\phi_{\vac k}\psi_{\vac k}=\frac{1}{\sqrt V}\sum_{\vac k}\tilde\phi_{\vac k} e^{i\vac k\cdot\vac x},\quad \vac k=\frac{2\pi}{L}\vac n,\ \vac n\in\mathbb Z^d. 
\label{E-160}
\ee
The zero mode $\tilde\phi_0$ itself is worthy of being distinguished from other Fourier modes $\tilde\phi_{\vac k\not=0}$ and we write 
\bea
F_{\rm GLW}[\tilde\phi_0,\tilde\phi_{\vac k\not=0}]&\simeq& \half \Bigl(r
+\frac{3u}{2L^d}\sum_{\vac k\not=0}|\tilde\phi_{\vac k}|^2
\Bigr)\tilde\phi_0^2+\quarter\frac{u}{L^d}\tilde\phi_0^4-{h}{L^{d/2}}\tilde\phi_0\nnb\\&&\quad+
\half\sum_{\vac k\not=0}(r+|\vac k|^2)|\tilde\phi_{\vac k}|^2+\dots.\label{E162}
\eea
Here we have suppressed terms of higher order than Gaussian for the non-zero modes because only the zero mode contributes to the non-vanishing average order parameter below the critical temperature.
The non-zero modes do not manifest non-vanishing average values even in the ordered phase (see Section~\ref{SecNotations}). 
Therefore we have two terms corresponding to two types of Fourier mode. 
The DIV $u$ enters alongside the temperature field $r$  to couple with the quadratic term in the zero mode expansion but the non-zero modes do not suffer from this shift.
This means that, while the zero mode $\tilde\phi_0$ is governed by the DIV-adjusted  GFP, the non-zero modes $\tilde\phi_{\vac k\not=0}$ are only controlled by the GFP itself.

We refer to the anomalous scaling emanating from DIVs as manifested in equations~(\ref{EQ_95}) or (\ref{EQ_97}) as Q scaling.
We discuss this in more detail in Section~\ref{sec7}.
In Ref.~\cite{PhysRevE.90.062137},   Wittmann and Young analyzed non-zero modes  and concluded  {\em standard} FSS  holds there. 
Thus one has for  $\vac k\not=0$,
\be
\chi_{\vac k\not=0}
=L^d\langle|\tilde\phi_{\vac k\not=0}|^2\rangle_L
\sim L^2 . \label{E163}
\ee

This statement, however, hides another subtlety in the story of Q. 
Equation~(\ref{E163}) as {\em standard} FSS   can either refer to Landau scaling, or to the GFP scaling (see Table~\ref{tab6}) because both deliver an exponent $2$.
The standard picture is (or was) that  {\em standard} FSS refers to the Landau picture. 
Indeed, in our early contributions to this topic \cite{BERCHE2012115}, although we referred to the $\chi_L\sim L^2$ behaviour as {\em Gaussian} scaling, we had Landau or  MFT in mind. 
In a later publication, it became clear that it is Gaussian and not Landau scaling that is at play in the non-zero modes~\cite{PhysRevLett.116.115701}. 
To disentangle them we looked at the FSS behaviour of the non-zero modes for magnetisation as well. 
There, Gaussian FSS predicts
\be
m_{\vac k\not=0}=
\langle|\tilde\phi_{\vac k\not=0}|\rangle_L
\sim L^{-\frac {d-2}2}\label{E164}
\ee
while Landau FSS would deliver  $L^{-1}$.
In Ref.~\cite{PhysRevLett.116.115701}, Eq. (\ref{E164}) was shown to indeed be correct.
Nowadays we refer to the GFP scaling described in Section~\ref{Gscaling}  as G scaling to distinguish it from the Q scaling. 
It is G-scaling (not Landau) that manifests as Eqs~(\ref{E163}) and (\ref{E164}), associated with non-zero modes.
We keep the term standard FSS for that which comes from the Landau picture in Section~\ref{LMFT} even though it has no direct physical manifestation.

\section{Corrections to  finite-size scaling and their dominance at short distances}
\label{sec6.4}

\index{correction!to finite-size scaling}
 Luijten and Bl\"ote considered RG equations for the scaling fields in the $O(N)$ $\phi^4$ model and we extended to the $\phi^n$ model  (\ref{eq-36}) in Ref.~[\refcite{ourscipost}].
 The outcome is that the DIV contaminates not only the amplitudes but the manner in which the temperature and magnetic field enter the homogeneity assumption as well. 
 The RG equations 
\be
\frac{dr}{d\ln b}=   {y_t} r+
	p u,\quad
\frac{du}{d\ln b}=y_u u, \quad 
\frac{dh}{d\ln b}=   {y_h} h.\label{eq115}
\ee
lead to  free energy density taking the form
\be
f_L(  \tau,h,u)=L^{-d}{\mathscr F}^\pm[
L^{y_ t^*}(  \tau u^{-2/n}-\tilde p u^{(n-2)/n}L^{y_u-   {y_t}}),
L^{y_h^*}hu^{-1/n}
]\label{E122}
\ee
where 
\be
\tilde p=-\frac{p}{(y_u-   {y_t})}
\label{clip}
\ee
and the temperature scaling field is
\be
  \tau=r+\tilde p u.
\ee

From Eq.(\ref{E122}), two types of FSS  for thermodynamic quantities emerge --- leading and corrected\footnote{For a more elaborate discussion on the possible role of corrections, see Refs.~[\refcite{PhHLuijten,ourscipost}].}. 
It also governs the leading FSS of the pseudocritical temperature. 

Denoting  the first argument by
\be
X=L^{y_  t^*}(  \tau u^{-2/n}-\tilde p u^{(n-2)/n}L^{y_u-   {y_t}}),
\label{E1611}
\ee
we identify a fixed value of it, say $X_0$, as that for which a given thermodynamic variable in the form of a derivative --- susceptibility, for example --- peaks.
This is the pseudocritical temperature $  \tau_L$, and we now have \be
  \tau_L=X_0u^{2/n}L^{-y_  t^*}+\tilde p uL^{-(y_t-y_u)}.\label{eq-133}
\ee
Because $ y_t^* = y_t-\frac{2}{n}y_u \le y_t-y_u$ above $d_{\rm uc}$, the first term dominates for large enough $L$, so that the shift exponent is $\lambda = y_t^*$ instead of $\lambda = y_t = 1/\nu$ in Eq.(\ref{shiftexp}). 
This recovers Eq. (\ref{E95}) for the 5D Ising model, for example.

In our opinion, {the analytic inclusion of corrections to scaling in} equation (\ref{E122}) is also a central result.
{The two-term structure was already proposed in BNPY for FBCs but not for PBCs.
While they proposed that the free energy scales as Eq.(\ref{BNPYL1}) (with $b=L$) for PBCs, 
\begin{quote}
``for other boundary conditions, where the system has a surface, it is probably necessary to use both $\tau L^{y_y^*} $ and $\tau L^{1/\nu}$ for a complete asymptotic description''~\cite{PhysRevB.31.1498}. \end{quote}
The second term in Eq.(\ref{eq-133}) when $n=4$  
corresponds to BNPY's proposal and extends it to other boundary conditions, including PBCs. }

 Luijten and Bl\"ote also derived two different decay modes for the spin-spin correlation function on this basis. 
 In Ref.~[\refcite{PhysRevB.56.8945}], they differentiated the finite-size free energy density wrt two {\emph{local}} magnetic fields placed at positions ${0}$ and $\vac{x}$, ``assuming that the finite-size behaviour is identical to the $\vac x$ dependence of $g$.''
 With this neat trick, they can account for DIVs through the free energy and incorporate them into the correlation function, without compromising the belief that the correlation length itself is anything other than linear in its dependency on finite size. 
 As they say [we set the long-range interaction-decay exponent to 2 for the short-range model under consideration here], 
 \begin{quote}
 ``If we do not take into account the dangerous irrelevant variable mechanism, we find 
 $g \propto L^{2y_h - 2d} 
 = 
 L^{-(d-2)}$, just as we found before from $\eta = 0$. 
 However, replacing $y_h$ by $y_h^*$ yields $g \propto L^{-d/2}$, in agreement with the $L$ dependence of the magnetic susceptibility. 
 This clarifies the difference between the two predictions: At short distances (large wave vectors),  there is no “dangerous” dependence on $u$. Hence, the finite-size behaviour of the spin-spin correlation function will be given by $L^{-(d-2+\eta)}$. For $ k = 0$, the coefficient of the $\phi^2$ term vanishes and thus the $u  \phi^4$ term is required
 which implies that $y_h$ is replaced by $y_h^*$ and $g_L$ scales as $L^{2y_h^*-2d}$.''
 \end{quote}
Thus Luijten and Bl\"ote arrived at two modes for the correlation function, the relative magnitudes of $r$ and $L$ determining which one applies. 
For short distances G modes dominate and for long distances Q modes reign.

More explicitly, taking corrections into account, for Q modes they have\footnote{Of course Luijten and Bl\"ote don't use the notation $g_G$ for G-modes and $g_Q$ for Q-mode, so we are somewhat preemptive in introducing our notation at this point.}
 \be
   g_Q(L/2) = L^{2y_h^*-2d}[c_o 
                        + c_1  \hat{t}   L^{y_t^*}  
                        + c_2  \hat{t}^2 L^{2y_t^*}   + \dots]
  \ee
  with $\hat{t}\equiv \tau+{\rm Const} \ \! L^{y_u-y_t}$.
They have a counterpart formula for log corrections.
For G modes, on the other hand, they have
 \be
   g_G(L/2) = L^{2y_h -2d}[c_o 
                        + c_1  \hat{t}   L^{y_t}  
                        + c_2  \hat{t}^2 L^{2y_t}   + \dots].
  \ee
There are no logarithms at the upper critical dimension for G modes because G modes are not affected by dangerous irrelevancy.

\begin{figure}[ht]
\centering
\includegraphics[width=0.70\textwidth]{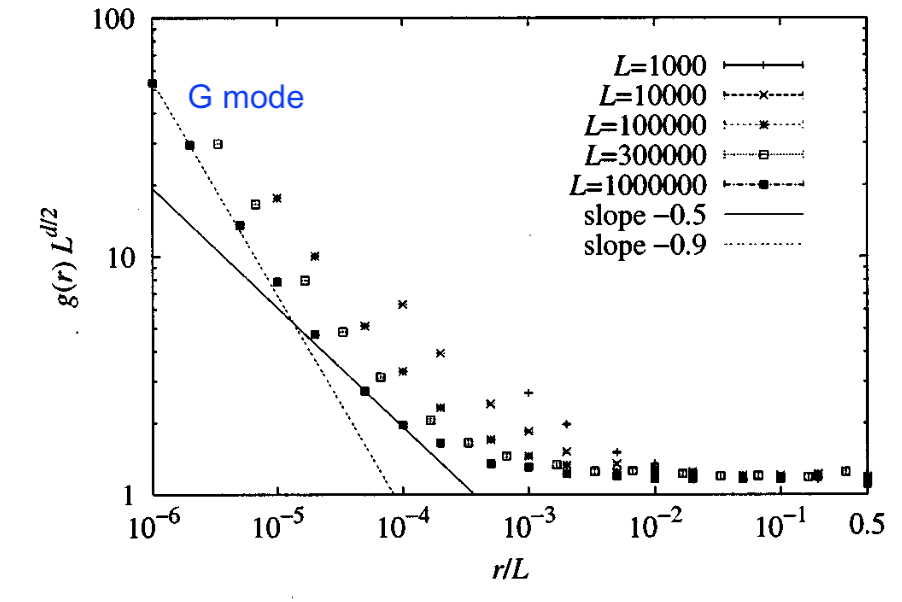}
\caption{The spin-spin correlation function versus $r/L$ in the 1D model with $\sigma=0.1$. Figure from E. Luijten and H.W.J. Bl\"ote, Classical critical behaviour of spin models with long-range interactions,  Phys. Rev. B {\bf 56}, 8945--8958, 1997,   \href{https://link.aps.org/doi/10.1103/PhysRevB.56.8945}{10.1103/PhysRevB.56.8945}.}
\label{fig:LB-Correl}
\end{figure}

Thus we have that to leading order, and above the critical dimension, 
$g_Q \sim g_G \times L^{-2y_u/n}$ and, since $y_u = - 2(d-d_{\rm uc})/(d_{\rm uc}-2)$ is negative there, $g_Q$ decays slower than  $g_G$. 
To determine if the two-decay-mode structure is visible in finite systems they simulated the long-range Ising model in one dimension and found that, indeed it is --- provided the system size is large enough to leave enough space to see it. 
Since this observation has been overlooked in recent literature \cite{Deng1} we reproduce it here (Fig.~\ref{fig:LB-Correl}).

In terms of critical exponents, the more traditional way of presenting the correlation function is through the scaling dimension as in Eq.(\ref{eq-11}) or the anomalous dimension as in Eq.(\ref{eq-5}).
In these terms, the Q-mode decay at $\tau=0$ is
 \be
   g(\vac x) \sim \frac{1}{|\vac x|^{2d-2y_h^*}} 
         =   \frac{1}{|\vac x|^{2x_\phi^*}}
         =   \frac{1}{|\vac x|^{d-2 + \eta^*}}
         =   \frac{1}{|\vac x|^{d-2\qqq}}
         =   \frac{1}{|\vac x|^{2d/n}}
  \ee
  with $\eta^*=d+2-2y_h^* = 2-2\qq$,
while the G-mode decay is
 \be
   g(\vac x) \sim \frac{1}{|\vac x|^{2d-2y_h}} 
         =   \frac{1}{|\vac x|^{2x_\phi}}
         =   \frac{1}{|\vac x|^{d-2 + \eta}}
         =   \frac{1}{|\vac x|^{d-2}}.
  \ee
  Quoting Luijten and Bl\"ote again, 
\begin{quote}``The fact that the $L$ dependence of $g(L/2)$ is determined by the $k=0$ mode raises the question of whether one can also observe the power-law decay described by $\eta$ [$=0$] in finite systems.''\end{quote}
To this end, they sampled the correlation decay as a function of $\vac x$ far into the high-dimensional regime.
They found that very large lattice sizes are required to leave enough space to observe the short-distance, $\eta=0$, G-mode decay at the start (for small $\vac x$ relative to large $L$). 
As stated by Luijten and Bl\"ote, for $\vac x$ of the order of the system size, \begin{quote}``the correlation function levels off. 
This is the mean-field contribution to the correlation function which dominates in the spatial integral yielding the magnetic susceptibility.''\end{quote}

In an attempt to fit it into the Q-scheme yet another approach was introduced in Ref.~[\refcite{KB2014}].
The standard derivation of Fisher's fluctuation-response relation (\ref{oldeta}) involves integrating the correlation function (\ref{defNu}) over the full infinite-size correlation volume $\xi_\infty^d$ to obtain the susceptibility close to the critical point. 
The same scaling relation ensues for finite volume provided the correlation length $\xi_L$ is bounded by the length. 
However, we now know this is not the case.
According to this view, the integral limits should depend upon whether one integrates over a scale of length $L$ and dimension $d$ or correlation length $\xi_L$ and dimension $d_{\rm uc}$.  
This delivers two different anomalous dimensions --- $\eta = 0$ on the scale of $\xi_L$ and $\eta_Q=2- \q \gamma / \nu$ on the scale of $L$. \index{anomalous dimension}
Thus the standard version of Fisher's scaling relation  (\ref{oldeta}) holds in high dimensions if length-scale is taken as correlation length. 
A corresponding counterpart to the fluctuation-dissipation relation (\ref{defNu}) ensues and is
\be
 \eta_Q = \q \eta + 2 (1-\q).
\ee

The issue of correlation decay in high dimensions had previously been addressed by John Nagle and Jill Bonner in 1970 \cite{Nagle_1970}.
In a numerical study, they found a negative value for the anomalous dimension of a one-dimensional chain with long-range interactions which took it above the upper criticality. They termed this $\tilde{\eta}$.
George Baker and Geoffrey Golner attempted to provide an explanation a few years later \cite{PhysRevLett.31.22}.
They suggested a difference between ``short long-range order'' and  ``long long-range order'' with only the former controlled by the standard anomalous dimension.  The negative anomalous dimensions of Nagle, Bonner, Luijten and Bl\"ote come from measuring distance on a scale of $\xi_L$.
Thus the $\eta^*$ of Luijten and Bl\"ote is the same as the  $\tilde{\eta}$ of Nagle and Bonner.

Finite-size effects in the high-dimensional Ising model were investigated in an outstanding PhD thesis by Vassilios Papathanakos (under the supervision of Michael Aizenman) in 2006\cite{Papathanakos}.
\begin{quote}``These results are consonant with recent findings in the theories of percolation and loop-erased random walks, which contribute to an emerging picture of multi-scale criticality.''\end{quote}
The effect of boundary conditions was investigated using a random-geometric approach coupled with a rigorous study of the Ising model on a high-dimensional box of volume $L^d$. 
Papathanakos found that 
\begin{quote}
``the short-range behaviour of the model is essentially unaffected by the choice of boundary conditions, and can be understood in terms of non-intersecting, independent simple random walks; 
this is reflected in the behaviour of bulk quantities, at temperatures where the correlation
length is small enough. 
However, as the critical temperature is approached, the probability for the intersection of the paths in the case of periodic boundary conditions becomes significant: the reentrance temperature sets the scale for the critical susceptibility, which is found to be at least of order
[$\xi_L(T_c) \gtrsim L^{d/2}$] in contrast to the case of bulk boundary conditions, where [$\xi_L(T_c) \lesssim  L^{d/2}$].

Moreover, the periodic two-point function at criticality can be essentially decomposed as the sum of the infinite-lattice two-point function, and a mean-field plateau, which dominates in the
bulk. 
Furthermore, making the assumption that the susceptibility is not larger than [$O(L^{d/2})$], we can also show that the renormalized coupling constant $g_L(T_c)$ of the system with periodic boundary conditions has a non-vanishing scaling limit, unlike the case of bulk boundary conditions, where the scaling limit is Gaussian. This assumption also implies that the value of the plateau at criticality is fixed at
[$g_L = O(L^{-1/2})$].
These findings are in essential agreement with recent numerical results present in the literature, and forward a more clear and intuitive explanation than theoretical estimates based on renormalization group arguments.''
\end{quote}

The main results for the Ising model are that the FSS for the susceptibility with FBCs is $\chi_L(T_c) \sim L^2$ while that for PBCs is 
$\chi_L(T_c) \sim L^{d/2}$.
Also for PBCs, the correlation function has reentrant behaviour vis
\be
 g_L(T_c,\vac x) \sim 
    \begin{cases}
      \frac{C_1}{|\vac x|^{d-2}} & \text{if } |\vac x| \le C_3 L^{\frac{d}{2(d-2)}}\\
      \frac{C_2}{|\vac x|^{1/2}} & \text{if } |\vac x| \ge C_3 L^{\frac{d}{2(d-2)}}.
    \end{cases}
\ee

This brings us to the close of the century.
But this by no means closes the story. 
In 2000 Binder and Luijten say \cite{problem}:
\begin{quote}
``Another issue where there has been a longstanding controversy between theory and simulation is the question of finite size scaling at dimensionalities $d$ above the marginal dimension $d^*$ where mean field theory becomes valid [22–30]. For systems with short-range forces, $d^*=4$, so in $d = 5$ all critical exponents (including corrections to scaling) are known, and hence finite-size scaling methods can be exposed to a stringent test. However, it will be shown that the comparison between theory [29] and simulation [30] is still disappointing!''
\end{quote}

They conclude with \cite{problem}: 
\begin{quote}
“In this work, two simple aspects of the bulk critical behaviour of ferromagnetic Ising models were
discussed, and it was shown that both problems (crossover from one universality class to another, and finite-size scaling above the marginal dimension) still are incompletely understood.”
\end{quote}
%
%

\section{Extension of dangerous irrelevancy in the correlation sector: the use of \coppa, the rescue of hyperscaling and finite-size scaling}
\label{sec7}

\index{dangerous irrelevant variable}
Forty years ago Fisher rescued the RG framework for the free-energy sector in the thermodynamic limit (Sec.\ref{breakthrough}) above the upper critical dimension, via his introduction of DIVs which were then extended to finite-size scaling 
corrections (Sec.\ref{sec6.4}).
However, problems still remain and, following Ref.~\cite{ourscipost},
we further build on Fisher's legacy to preserve the RG framework,
 extending DIVs to the correlation sector.
{In doing so,  we closely follow} Luijten and Bl\"ote's approach that we  proposed to generalize  in Ref.~\cite{PhysRevLett.116.115701}


We return firstly to Section~\ref{DIVFSS} and BNPY's extension of the DIV mechanism to the free energy itself.
As stated, they assumed that the correlation length could obey a similar homogeneity law (\ref{eq-104}) but insisted that $q_1=0$ and left $q_2$ and $q_3$ open. 
For the free energy in Eq.(\ref{eq-102}), BNPY gave three reasons why $p_1=0$ and hence $d^*=d$. 
In a series of papers, e.g. in Ref.~[\refcite{CMP2013}], we advocated that the correlation length should be allowed to exceed the length --- as Br\'{e}zin had shown for $d=4$ and general $n$.
In particular, we showed that 
\be
 \xi_L \sim L^{\qqq} 
\ee
with 
\be
\mathcoppa=d/d_{\rm uc}.
\label{coppadoverdc}
\ee
Therefore we write
\be
\xi_L(t,h,u)=b^{\qqq}\Xi^\pm(b^{y_t^{+}}tu^{q_2},b^{y_h^{+}}hu^{q_3}).
\label{Fri1}
\ee
where 
\be 
\mathcoppa= 1+q_1y_y,\quad
y_ t^+ =   {y_t}+ q_2 y_u,\quad y_h^+ =   {y_h}+q_3 y_u\, .
\label{Fri2}
\ee
The value for $\mathcoppa$ in Eq.(\ref{coppadoverdc}) determines $q_1=-1/n$.
In the thermal sector, matching Eq.(\ref{Fri1}) with the value $\nu_{\rm\scriptscriptstyle MFT} = 1/2$ in Eq.(\ref{eq-59}) determines 
$q_2=-2/n$, identical to $p_2$. 
Likewise, matching Eq.(\ref{Fri1}) with the value $\nu_{\rm\scriptscriptstyle c\ \!MFT} = (n-2)/2(n-1)$ in Eq.(\ref{eq-59})  determines 
$q_3=-1/n$, identical to $p_3$.
Note that, while the correct correlation-length critical exponent $\nu_{\rm\scriptscriptstyle MFT}$ is identical to that coming from the GFP 
$\nu_{\rm\scriptscriptstyle GFP} = 1/y_t=1/2$ in Eq.(\ref{defNu}), their counterparts in the magnetic sector do not coincide and
$\nu_{\rm\scriptscriptstyle c\ \!MFT}$ differs from
$\nu_{\rm\scriptscriptstyle c\ \!GFP} = 1/y_h$ from Eq.(\ref{defNub}).
So, technically although $\nu_{\rm\scriptscriptstyle MFT} = 1/y_t$, this is by coincidence only and the correct expression is $\nu_{\rm\scriptscriptstyle MFT} = \qq/y_t^+ = \qq/y_t^*$. 
Note also that because $q_2=p_2$ and $q_3=p_3$, one has $y_t^+=y_t^*$ and $y_h^+=y_h^*$. We henceforth only use the starred scaling dimensions.
I.e., to extend Eqs.(\ref{SH1}) and (\ref{SH2}) to the correlation sector, we write
\bea
&\displaystyle \nu_{\rm\scriptscriptstyle MFT}=\frac{   {{\qq}}}{   {{y_t^*}}} \qquad
&\nu_{\rm\scriptscriptstyle c\ \!MFT}=\frac{\qq}{   {{y_h^*}}}.
\label{SH3}
\eea
Equations (\ref{E1041}) and (\ref{E1041b}) can now be written above $d_{\rm uc}$ in a very consistent manner as
\be
d^*=\qq d_{\rm uc},\quad y_t^*(d)=\qq y_t(d_{\rm uc}),\quad y_h^*(d)=\qq y_h(d_{\rm uc}).
\ee

Likewise, we believe that there is no reason to invoke a mechanism by which the rescaling of the temperature and the magnetic field would differ for different physical quantities. Therefore, the result that  $y_t^+=y_t^*$ and $y_h^+=y_h^*$ is satisfactory. A second extension proposed is to use the correct (starred) scaling dimension ($2x_\phi^*$ in fact) 
of the matter field to describe the dimension of the correlation function. 
From these considerations, we are led to the following scaling hypotheses (equation (\ref{E122}) is rewritten for the sake of clarity),
\bea&&f_L(  \tau,h,u)=b^{-d}{\mathscr F}^\pm(X,Y,bL^{-1}),\label{eq-134}\\
&&g_L(\vac x,  \tau,h,u)=b^{-2x_\phi^*}{\mathscr G}^\pm(b^{-1}\vac x,
X,Y,bL^{-1}),\label{eq-135}\\
&&\xi_L(  \tau,h,u)=b^{\qqq}\Xi^\pm(X,Y,bL^{-1}).\label{eq-136}
\eea
These are expressed in terms of the two rescaled variables
\bea
&&X=b^{y_  t^*}(  \tau u^{-2/n}-\tilde p u^{(n-2)/n}b^{y_u-   {y_t}}),\label{eq-137}\\
&&Y=b^{y_h^*}hu^{-1/n},\label{eq-138}\eea
and with the scaling dimensions
\be
y_  t^*=d\Bigl(1-\frac 2n\Bigr),\ y_h^*=d\Bigl(1-\frac 1n\Bigr),\ x_\phi^*=\frac dn,\ \mathcoppa=
\frac d2\Bigl(1-\frac 2n\Bigr)\label{eq-139}
\ee
having used 
(\ref{E1041}) with (\ref{eq-62}), (\ref{eq-63}), (\ref{eq-64}) and extended $x_\phi^*$ from Ref.~\cite{PhysRevLett.116.115701} for the $n=4$ case, and where
$\tilde p$ is given in Eq.(\ref{clip}).
{Obviously these scaling forms apply to the  Q-sector only (Fourier $Q$-modes). 
The G sector is unaffected by DIVs and the finite-size counterparts of Eqs.(\ref{eq-10}), (\ref{eq-11}) and (\ref{eq-12}) apply there.  }
Following a suggestion by Michael Fisher, a new exponent  $\mathcoppa$  was introduced in Ref.~[\refcite{CMP2013}] {for the  Q sector}. We like the use of this exponent, since it is very easy to translate equations from one universality class to another in terms of $d/d_{\rm uc}$, and also to generalize from the GFP values (which we recover reverting $\qq$ to 1), so one can also rewrite the exponents (\ref{eq-139}) in the form
\be
y_t^*=2\mathcoppa,\ y_h^*=\frac d2+\mathcoppa,\ x_\phi^*=\frac d2-\mathcoppa,\ \mathcoppa=\frac d{d_{\rm uc}}.
\ee

An interesting use of $\mathcoppa$ is in a new form of the hyperscaling relation.
Setting $b=|\tau|^{-1/y_t}$ in (\ref{eq-136}) delivers $\xi_\infty\sim|\tau|^{-\qqq/y_t^*}$, hence $y_t^*=\mathcoppa/\nu$, then in (\ref{eq-134}) we get $f_\infty\sim|t|^{d/y_t^*}\sim|\tau|^{2-\alpha}$ which leads to 
\be
\alpha=2-\frac{\nu d}{\qqq}.
\label{hyperhyper}
\ee
This repairs\footnote{Since 
Josephson's inequality $\nu d \ge 2 - \alpha$ was introduced in 1967 \cite{Josephson67}, {literature, including} textbooks, on statistical physics, lattice field theory, etc. refer to  hyperscaling as ``failing" above the upper critical dimension.  
This statement should no longer be used in statistical physics --- hyperscaling does not fail because the RG does not fail above the upper critical dimension. 
{Moreover, the}  hyperscaling relation should rather be rewritten properly {as (\ref{hyperhyper}) and not as (\ref{oldhyper}).} } the hyperscaling relation above $d_{\rm uc}$, and it holds also below $d_{\rm uc}$ where $\mathcoppa=1$.
{(Obviously $\mathcoppa=1$ for the non-zero or orthogonal Fourier modes in high dimensions too.) }

The new exponent already enabled predictions to be more naturally expressed  for models such as the nearest-neighbour Ising model, {percolation} above its critical dimension $d_{\rm uc}=6$  and for Long-Range Ising Models (LRIM' with various dimensions above $d_{\rm uc}(\sigma)$  (with periodic boundary conditions)
\cite{lotsfq,Deger3}.
The extension to general values of $n$ is obvious and  we collect the predictions for arbitrary $d>d_{\rm uc}$ for quantities which have been discussed above:
\bea
&&m_L\sim L^{-(\frac d2-\qqq)},\ \chi_L\sim L^{2\qqq},\ e_L\sim L^{-(d-2\qqq)},\ c_L\sim L^{4\qqq-d},\label{eq-140}\\
&&\xi_L\sim L^{\qqq},\ g_L(X_0)\sim L^{-(d-2\qqq)},\label{eq-141}\\
&&  t_L\sim L^{-2\qqq},\ \Delta\beta_L\sim L^{-2\qqq},\ |h^L|\sim L^{-(\frac d2+\qqq)},\\
&&h^{\rm LY}_L\sim L^{-(\frac d2+\qqq)}, \ t^{\rm F}_L\sim L^{-2\qqq}.\label{eq-202}
\eea
Note that all of the above scaling formulas are readily obtained by replacing the standard FSS prescription
 Eq.(\ref{eq87}) by the prescription
\be Q_\infty(  t,0)\sim |  t|^\rho
\longrightarrow 
 Q_L(  t=0,0)\sim L^{-\qqq \rho/\nu}.
 \label{QFSS}
 \ee
 This is what we call Q-scaling and it holds at the pseudocritical point as we shall see shortly.
 With Q-scaling to hand, FSS holds above the upper critical dimension.

\index{critical!dimension!upper}
\begin{table}[ht]
\tbl{Summary of the evolution of the scaling picture above the upper critical dimension for the $\phi^4$ model.
The first column presents the correct results (FSS predictions are for a system with periodic boundary conditions). 
In the other columns, we give the (incorrect) results predicted when they are different. 
A question mark means that the quantity hasn't been considered in the corresponding scenario.
 }
{\begin{tabular}{@{}llllll@{}} \toprule
The correct results & Landau scaling$^1$ & GFP$^2$ & Fisher DIV$^3$ & BNPY$^4$ & \coppa$^5$ \\
\hline \\
thermodynamic limit:\\
$c_\infty(  \tau,0)\sim|  \tau|^0$ & $\surd$ & $\alpha_{\rm\scriptscriptstyle G}=2-\frac d2$ &  $\surd$ & $\surd$ & $\surd$ \\
$m_\infty(  \tau,0)\sim|  \tau|^{1/2}$ & $\surd$ & $\beta_{\rm\scriptscriptstyle G}=\frac{d-2}4$ &  $\surd$ & $\surd$ & $\surd$  \\
$\chi_\infty(  \tau,0)\sim|  \tau|^{-1}$ & $\surd$ & $\surd$ &  $\surd$ & $\surd$ & $\surd$ \\
$m_\infty(0,h)\sim|h|^{1/3}$ & $\surd$ & $\delta_{\rm\scriptscriptstyle G}= \frac{d+2}{d-2}$ &  $\surd$ & $\surd$ & $\surd$ \\
$\xi_\infty(  \tau,0)\sim|  \tau|^{-1/2}$ & $\surd$ & $\surd$ &  $\surd$ & ? & $\surd$ \\
$g_\infty(\vac x,0,0)\sim|\vac x|^{-(d-2)}$ & $\surd$ & $\surd$ &  $\surd$ & ? & $\surd$ \\ \\
\hline \\
FSS:\\
$\xi_L(\tau=0,0)\sim L^{d/4}$
& $L$ & $L$ & $ L$ & $L$ & $\surd$\\
$\chi_L(\tau=0,0)\sim L^{d/2}$
 & $L^2$ & $L^2$ & $L^2$ & $\surd$ & $\surd$\\
$  \tau_L(\tau=0,0)\sim L^{-d/2}$
 & $L^{-1/2}$ & $L^{-1/2}$ & $L^{-1/2}$ & $\surd$ & $\surd$\\
$m_L(\tau=0,0)\sim L^{-d/4}$ & $L^{-1}$ & $L^{1-d/2}$ & $L^{-1}$ & $\surd$ & $\surd$\\
$g_L(L/2,\tau=0,0)\sim L^{-d/2}$ & $L^{-(d-2)}$ & $L^{-(d-2)}$ & $L^{-(d-2)}$ & ? & $\surd$ \\
$h_L^{\rm LY}(\tau=0)\sim L^{-3d/4}$ & $L^{-3}$ & $L^{-(d+2)/2}$ & & & $\surd$ \\
$  \tau_L^{\rm F}(\tau=0)\sim L^{-d/2}$ & $L^{-2}$ & $L^{-2}$ & & & $\surd$
\\
\\\botrule
\end{tabular}}
\begin{tabnote}
 $^1$FSS with Landau exponents, 
 
 $^2$Predictions from  the RG eigenvalues at the Gaussian Fixed Point, 
 
 $^3$Corrections made by the scenario of Fisher,  
 
 $^4$Most of the results presented in this column correspond to BNPY's version of the scenario of Fisher and are checked in Ref.~\cite{Binder85}, 
 
 $^5$Q scaling
 of the results presented in the last column are checked e.g. in Ref.~\cite{BERCHE2012115}.

\end{tabnote} \label{tab6}
\end{table}

 The different approaches that we have discussed are collected and compared in Table~\ref{tab6} for the $\phi^4$ model with periodic boundary conditions. 
 In this table, the first column lists known results in the thermodynamic limit or for finite-size scaling. 
 The remaining columns correspond to the different approaches that we have described so far, with the symbol $\surd$ to denote an agreement and, {in cases of disagreement}, the prediction made by the (incorrect) theory considered is given explicitly. The last column is for Q scaling.

\section{The case of Free Boundary Conditions}\label{sec9}
\subsection{A mismatch between $T_L$ and $T_c$ in free boundary conditions}
The case of free boundary conditions is more problematic. 
First the easy part: the study of FSS properties of physical quantities evaluated {\em at the pseudo-critical point} for FBCs agrees with what we said previously for systems with PBCs. 
What is happening right at the critical point on the other hand is still subject to debate.

Lundow and Markstr\"om have studied this problem intensively for the Ising model in the 5D case, for very large systems (up to $L=160$ in Ref.~[\refcite{LUNDOW2014249}]). They obtained, for the magnetisation and the susceptibility, the leading behaviours and corrections to scaling of the form:
\bea
&&\chi_L(T_c)=0.817\ \!L^2+0.083\ \!L,\label{E177}\\
&&m_L(T_c)=0.230\ \!L^{-3/2}+1.101\ \!L^{-5/2}-1.63\ \!L^{-7/2}.\label{E187}
\eea
There, the leading and correction exponents are fixed and the coefficients are free, and the authors concluded in favour of a { standard FSS behaviour} of the susceptibility at the critical temperature  $\chi_L(T_c)\sim L^2$. The same conclusion, albeit with much lower accuracy, was reached in  \cite{BERCHE2012115}. There, by {\em standard}  it was referred to FSS with Landau exponents, $\chi_L\sim L^{\gamma_{\rm\scriptscriptstyle MFT}/\nu_{\rm\scriptscriptstyle MFT}}$, a conclusion that we will see is { not} correct, although $L^2$ is (accidentally) correct.

The case of the magnetisation in (\ref{E187}) is a bit more subtle, because the leading exponent $-\frac 32$ { is not} the standard Landau FSS (the ratio of MFT exponents would be $-\beta_{\rm\scriptscriptstyle MFT}/\nu_{\rm\scriptscriptstyle MFT}=-1$ instead).

In order to clarify the situation, Wittmann and Young~\cite{PhysRevE.90.062137,WittmannThesis}, and then Flores-Sola et al.~\cite{PhysRevLett.116.115701}
 considered the behaviour of the Fourier modes in a finite system with free boundary conditions.
Following Ref.~\cite{PhysRevB.32.7594}, 
a sine-expansion of the scalar field in the $\phi^4$ action is performed with the boundary conditions 
$\phi({\bf x})=0$ at the free surfaces. 

Wittmann and Young confirmed the scaling at $T_c$ for the total susceptibility (for FBC). They found that the single-mode susceptibility also obeys a
{ standard FSS behaviour}, $\chi_{\vac k}\sim L^2$ for the modes
which { will not acquire a nonzero magnetisation}, namely with the smallest wave-vector with an even wave number $n_\alpha$ in the wave expansion ($k_\alpha=n_\alpha \pi/(L+1)$, $n_\alpha=1,2,\dots,L$, $\alpha=1,\dots\ ,d$). Wittmann and Young proposed to use $ t=T-T_L$, with $t=\tau+{\rm const}\ \!\times  L^{-\lambda}$,
 as the temperature-like scaling variable and the starred RG dimensions in the scaling hypothesis for the susceptibility. They obtain then 
$\chi_L( t)=L^{2y_h^*-d}{\mathscr X}(L^{y_t^*} t)$ in zero magnetic field and  the compatibility with 
the bulk behaviour in the thermodynamic limit
$\chi_\infty(\tau)\sim|\tau|^{-\gamma_{\rm\scriptscriptstyle MFT}}$ is recovered  by the demand that the asymptotic regime obeys ${\mathscr X}(x)\sim x^{-\gamma_{\rm\scriptscriptstyle MFT}}$ (with $\gamma_{\rm\scriptscriptstyle MFT}=1$ here) for $x\to \infty$.
Thid leads to  $\chi_L( t)\sim L^{2y_h^*-d+y_t^*} |t|^{-1}$.
When $\lambda=d/2$, as is the case in PBC, this leads at criticality to the correct result $\chi_L(T_c)=L^{d/2}$. On the other hand if $\lambda =2$ (FBC), this amounts to
$\chi_L(T_c)=L^{2}$, hence the scenario proposed is very compelling.

The discussion concerning standard FSS with Landau exponents, or Gaussian FP exponents is nevertheless not settled by the study of the susceptibility alone and here we will present additional unpublished numerical results~\cite{flore2016} for low-dimensional Ising models with long-range interactions above their respective upper critical dimensions.

\subsection{The three possible scenarios}

As we have already noted several times, one cannot disentangle, with the susceptibility alone, the predictions of { standard} (or Landau) FSS from those of the { Gaussian Fixed point} FSS. 
The primary aim of Ref.~\cite{PhysRevLett.116.115701} was to test the discriminating case of the magnetisation.
The argument of Wittmann and Young for the magnetisation would indeed lead to $m_L(T_c)=L^{-\lambda/2}$, hence
$m_L(T_c)=L^{-d/4}$ in PBC and  $m_L(T_c)=L^{-1}$ in FBC.
The argument  is thus equivalent to { Landau scaling}, because for FBC, $\lambda=1/\nu_{\rm\scriptscriptstyle MFT}$.

So we are led to the point where one essentially faces three distinct hypotheses\footnote{In Ref.~\cite{ourscipost}, a fourth scenario has been considered in which corrections to scaling could be dominant at small sizes, but this hypothesis has not been confirmed by the numerics. Therefore we do not discuss this fourth case further here.}  among which one has to discriminate (see Table~\ref{table11}).

\begin{table}[ht]
\tbl{The three scenarios for the FSS of the susceptibility and the magnetisation at $T_c$. There, $\qq=d/d_{\rm uc}$, $d_{\rm uc}$ and the ${\rm MFT}$ exponents take their usual values and extension to different universality classes is obvious.}
{\begin{tabular}{@{}lll@{}} \toprule
Hypothesis & Scaling of $\chi_L(T_c)$ & Scaling of $m_L(T_c)$ \\
\hline 
1. Landau (or { standard}) scaling & $L^{\frac{\gamma_{\rm\scriptscriptstyle MFT}}{\nu_{\rm\scriptscriptstyle MFT}}}$   & $L^{-\frac{\beta_{\rm\scriptscriptstyle MFT}}{\nu_{\rm\scriptscriptstyle MFT}}}$ \\
2. Q scaling & $L^{2\qqq}$ & $L^{\qqq-d/2}$ \\
3. G scaling & $L^{2}$ & $L^{1-d/2}$ \\
\botrule
\end{tabular}}
\label{table11}
\end{table}

In the case of the $\phi^4$ universality class, the options are the following:
\begin{description}
\item{---}  prediction 1 (Landau scaling) leads to $\chi_L(T_c)\sim L^2$ and $m_L(T_c)\sim L^{-1}$, 
\item{---} prediction 2 (Q scaling) to $\chi_L(T_c)\sim L^{d/2}$ and $m_L(T_c)\sim L^{-d/4}$, 
\item{---} prediction 3 (G scaling)  to $\chi_L(T_c)\sim L^2$ and $m_L(T_c)\sim L^{-(d-2)/2}$.
\end{description}
Option 2 is clearly ruled out at $T_c$ by the results of Lundow and Markstr\"om in (\ref{E177}) (as well as by the results of Wittmann and Young). Options 1 and 2 are compatible with the results measured for the susceptibility. For the 5D model, it is easy to discriminate, with the magnetisation,  between option 1 (which predicts the value $-1$), option 2 (prediction $-1.25$) and option 3 (prediction $-1.5$). Equation (\ref{E187}) is in favour of option 3.

\subsection{Towards a Ga	ussian Fixed Point scaling at $T_c$}
\index{Gaussian!fixed point}
In Ref.~\cite{PhysRevLett.116.115701} the magnetisation of the LRIM was investigated, tuning the parameter $\sigma$ of the interaction decay to control the upper critical dimension $d/(2\sigma)$. 
In the thesis of Emilio Flores-Sola~\cite{flore2016} results are reported for various values of $\sigma $ and of $d$. 
Some of these are collected in Tables~\ref{tabLRIMMag}, \ref{tabLRIMXi} and \ref{tabLRIMCorrFunct} for the magnetisation, the susceptibility, correlation length and correlation function at $T_c$ and at $T_L$. 
In the case of the magnetisation (Table~\ref{tabLRIMMag}), the expectation from GFP scaling is an exponent $-\beta_\G/\nu_\G$ with $\beta_\G=(d-\sigma)/(2\sigma)$ and $\nu_\G=1/\sigma$, while Q scaling predicts an exponent $-\qq\beta_\MFT/\nu_\MFT=-\qq$.

\begin{table}[h!]
\tbl{FSS of the magnetisation (upper panel) and the susceptibility (lower panel) for the LRIM in $d=1$, 2  with FBCs at $T_c$ and at $T_L$.
The results at the critical temperature support G scaling ($\frac{d-\sigma}2$, resp. $\sigma$) while those at the pseudo-critical temperature support Q scaling ($d/4$, resp. $d/2$). }
{\begin{tabular}{@{}ll | cc | cc@{}} \toprule
&& $m_L(T_c)$ & $\frac{d-\sigma}2$ & $m_L(T_L)$ & $\frac {\qq \beta_\MFT}{\nu_\MFT}=\frac d{4}$\phantom{$\frac {\qq \beta_\MFT}{\nu_\MFT}$}\\
\hline
$d=1$ & $\sigma=0.1$ & $L^{-0.450(4)}$ & 0.45 & $L^{-0.233(4)}$ & 0.25\\
           & $\sigma=0.2$ & $ L^{-0.401(3)}$& 0.40 & $L^{-0.230(4)}$ & 0.25\\
\hline
$d=2$ & $\sigma=0.1$ & $ L^{-0.949(1)} $ & 0.95 & $L^{-0.501(2)}$ & 0.50 \\
           & $\sigma=0.2$ & $ L^{-0.897(1)}$ & 0.90 & $L^{-0.494(1)}$ & 0.50 \\ \hline
&& $\chi_L(T_c)$ & $\sigma$ & $\chi_L(T_L)$ & $\frac{\qq\gamma_\MFT}{\nu_\MFT}=\frac d2$\phantom{$\frac {\qq \beta_\MFT}{\nu_\MFT}$}\\
\hline
$d=1$ & $\sigma=0.1$ & $L^{0.099(1)} $ & 0.1 & $L^{0.522(3)}$ & 0.5\\
           & $\sigma=0.2$ & $ L^{0.200(1)}$& 0.2 & $L^{0.525(5)}$ & 0.5\\
\hline
$d=2$ & $\sigma=0.1$ & $ L^{0.094(2)} $ & 0.1 & $L^{0.985(2)}$ & 1. \\
           & $\sigma=0.2$ & $ L^{0.198(2)}$ & 0.2 & $L^{0.994(2)}$ & 1. \\
\botrule
\end{tabular}}
\label{tabLRIMMag}
\end{table}

\begin{table}[h!]
\tbl{FSS of the correlation length for the LRIM in $d=1$, 2  with FBCs at $T_c$ and at $T_L$.
The results at the critical temperature support G scaling ($1$) while those at the pseudo-critical temperature are more in favour of Q scaling ($d/(2\sigma)$). }
{\begin{tabular}{@{}ll | cc | cc@{}} \toprule
&& $\xi_L(T_c)$ & $1$ & $\xi_L(T_L)$ & $\qq=\frac{d}{2\sigma}$\phantom{$\frac {\qq \beta_\MFT}{\nu_\MFT}$}\\
\hline
$d=1$ & $\sigma=0.1$ & $L^{1.01(3)} $ & 1 & $L^{4.03(7)}$ & 5\\
           & $\sigma=0.2$ & $ L^{1.03(2)}$& 1 & $L^{2.21(4)}$ & 2.5\\
\hline
$d=2$ & $\sigma=0.1$ & $ L^{1.07(7)} $ & 1 & $L^{7.48(4)}$ & 10 \\
           & $\sigma=0.2$ & $ L^{0.95(6)}$ & 1 & $L^{3.97(4)}$ & 5 \\
\botrule
\end{tabular}}
\label{tabLRIMXi}
\end{table}

\begin{table}[h!]
\tbl{FSS of the correlation function at $\vac x|=L/2$ for the LRIM in $d=1$, 2  with FBCs at $T_c$ and at $T_L$.
The results at the critical temperature are more in favour of G scaling (${d-\sigma}$) while those at the pseudo-critical temperature support Q scaling ($d/2$). }
{\begin{tabular}{@{}ll | cc | cc@{}} \toprule
&& $g_L(T_c,\frac L2)$ & $d-2+\eta_\G$ & $g_L(T_L,\frac L2)$ & $d-2+\eta_\qqq$\phantom{$\frac {\qq \beta_\MFT}{\nu_\MFT}$}\\
\hline
$d=1$ & $\sigma=0.1$ & $L^{-0.86(6)} $ & 0.9 & $L^{-0.487(5)}$ & 0.5\\
           & $\sigma=0.2$ & $ L^{-0.83(6)}$& 0.8 & $L^{-0.483(6)}$ & 0.5\\
\hline
$d=2$ & $\sigma=0.1$ & $ L^{-2.07(9)} $ & 1.9 & $L^{-0.954(3)}$ & 1 \\
           & $\sigma=0.2$ & $ L^{-1.70(9)}$ & 1.8 & $L^{-0.974(3)}$ & 1 \\
\botrule
\end{tabular}}
\label{tabLRIMCorrFunct}
\end{table}

Other quantities (temperature shift, correlation length, correlation function) are also reported in Ref.~[\refcite{flore2016}] (and partially collected here) and, although perfectible,  all numerical results at $T_c$ support the option 3 above (and also confirm Q scaling, i.e. option 2, at $T_L$).
A recent work\cite{Honchar} has revisited FSS in the 5D Ising model (with short-range interactions) with FBCs and also confirms the present results.

The picture now is the following for FBCs: 
{\emph{Either}} we study physical quantities which are related to the {Q modes} (i.e. modes that have a non-zero projection on the average magnetisation) or those related to the G modes (modes that do not contribute to the average magnetisation).
In the first case, which is expected for FBC at $T_L$, the DIV has to be taken into account and Q scaling rules (option 2 in Table~\ref{table11}).
In the second case, which holds for FBC at $T_c$,  {$u$ is not dangerous} and the physics is controlled by the GFP (option 3 in Table~\ref{table11}).
The exponents there are Gaussian exponents collected in Table~\ref{tab2} {\it and not Landau exponents}.

\section{Conclusion}
We have reached a point where we believe that $\qq$ and $\hat \qq$ are necessary ingredients of FSS at and above the upper critical dimension. They allow for a natural, and simple, extension of the scaling hypothesis that governs the behaviour of thermodynamic averages and correlations 
in systems having a non-zero finite-size order parameter. This is the case at the pseudo-critical points $T_L$, and also at $T_c$ for systems with PBCs. There, the scaling hypothesis for the singular part of the free energy density, for the correlation length and for the order parameter correlation function can be written as follows
\bea
d>d_{\rm uc}\\
&&f(\tau,h,L^{-1})=b^{-d}f(b^{2\qqq}\tau,b^{d/2+\qqq}h,bL^{-1}),\nnb\\
&&\xi(\tau,h,L^{-1})=b^{\qqq}\xi(b^{2\qqq}\tau,b^{d/2+\qqq}h,bL^{-1}),\nnb\\
&&g(\tau,h,L^{-1},\vac x)=b^{-(d-2\qqq)}g(b^{2\qqq}\tau,b^{d/2+\qqq}h,bL^{-1},b^{-1}\vac x),\nnb\\
d=d_{\rm uc}\\
&&f_{\rm uc}(\tau,h,L^{-1})=b^{-d_{\rm uc}}f(b^{2}(\ln b)^{\hat y_t}\tau,b^{d_{\rm uc}/2+1}(\ln b)^{\hat y_h}h,bL^{-1}),\nnb\\
&&\xi_{\rm uc}(\tau,h,L^{-1})=b\ \!(\ln b)^{\hat\qqq}\xi(b^{2}(\ln b)^{\hat y_t}\tau,b^{d_{\rm uc}/2+1}(\ln b)^{\hat y_h}h,bL^{-1}),\nnb\\
&&g_{\rm uc}(\tau,h,L^{-1},\vac x)=b^{-(d-2)}g(b^{2}(\ln b)^{\hat y_t}\tau,b^{d_{\rm uc}/2+1}(\ln b)^{\hat y_h}h,bL^{-1},b^{-1}\vac x)\nnb
\eea
and the scaling of other quantities follows from appropriate derivatives wrt the scaling fields.

Hyperscaling relation among critical exponents and ``hatted'' critical exponents of the logarithmic corrections take the form \index{hyperscaling relation}
\bea
&&\alpha=2-\frac{\nu d}{\qq},\\
&&\hat\nu=\hat\qq-\frac{\nu\hat\alpha}{2-\alpha}.
\eea
They are valid above, below, and at $d_{\rm uc}$ for the first one (with $\qq=1$ below and $\qq=d/d_{\rm uc}$ above $d_{\rm uc}$), and the second one is valid at $d_{\rm uc}$.

In FBCs, a special situation arises due to the rounding which does not scale like the shift of the pseudo-critical temperature, and, as a consequence, the pseudo-critical temperature is somehow pushed far below the critical temperature. It follows that at the critical temperature, there is no room for the order parameter to develop in the free energy mode expansion (it develops at the pseudo-critical point). The non-zero order parameter (this would be the zero-mode in the PBC case), which is responsible for the contamination of the scaling by the dangerous irrelevant variable, has no influence there. It follows that the correct homogeneity assumption that describes the singular behaviour in the very vicinity of the critical temperature $\tau\to 0$ is as above except that $\qq$ stays stuck to 1 and the physical quantities are controlled by the Gaussian fixed point. As far as we know, this is the only instance (FBC, $T=T_c$) in which ratios of the Gaussian exponents are measured in FSS and not ratios of MFT exponents. In a similar manner, one may expect that the logarithmic corrections in the Q sector which are due to the scaling field $u$ becoming marginal at $d_{\rm uc}$, do not show up in the G sector.

\section*{Acknowledgments}
We would like to thank Yu. Holovatch for his constant support, and for accepting a delay of about a year in the delivery of the present chapter!

This may not be a common practice, but here I would like also to thank two of my collaborators, two PhD students who are working with me, and were under the co-supervision of Ralph. Andy Manapany and Le\"\i la Moueddene. They are wonderful people with whom we can share opinions and views on the world. I think that maybe I opened their minds to some aspects, and I am sure that they opened mine. They knew Ralph, they appreciated him a lot, and we share this.

\bibliographystyle{ws-rv-van}
\bibliography{references}

\printindex                        

\end{document}